\documentclass[aps,rmp,reprint,amsmath,amssymb,longbibliography,floatfix]{revtex4-1}

\newcommand{\ket}[1]{\ensuremath{\left| #1 \right>}}

\newcommand{\braket}[1]{\ensuremath{\left< #1 \right>}}

\newcommand{\Rb}{\ensuremath{^{87}\text{Rb }}}

\newcommand{\pd}[2]{\ensuremath{\frac{\partial #1}{\partial #2}}}

\newcommand{\ie}{\emph{i.e.} }
\newcommand{\eg}{\emph{e.g.} }

\newcommand{\h}{\hbar}

\usepackage{graphicx}
\usepackage{epstopdf}
\usepackage{setspace}
\usepackage{bbold}
\usepackage{comment}
\usepackage{color}
\usepackage{soul}
\usepackage[abs]{overpic}
\usepackage{lipsum}
\usepackage{nicefrac}
\usepackage{mathtools}
\usepackage{amsthm}
\usepackage[colorlinks=true,citecolor=blue,linkcolor=blue]{hyperref}
\usepackage{bookmark}
\definecolor{fixmecolor}{rgb}{0.9290, 0.6940, 0.1250}

\begin{document}

\title{Anomalous statistics of laser-cooled atoms in dissipative optical lattices}

\author{Gadi Afek}
\affiliation{Quantum Art LTD, Rehovot 7414003, Israel}
\author{Nir Davidson}
\affiliation{Department of Physics of Complex Systems, Weizmann Institute of Science, Rehovot 76100, Israel}
\author{David A. Kessler} 
\affiliation{Department of Physics, Institute of Nanotechnology and Advanced Materials, Bar-Ilan University, Ramat-Gan 52900, Israel}
\author{Eli Barkai}
\affiliation{Department of Physics, Institute of Nanotechnology and Advanced Materials, Bar-Ilan University, Ramat-Gan 52900, Israel}

\date{\today{}}

\begin{abstract}
Diffusion occurs in numerous physical systems throughout nature, drawing its generality from the universality of the central limit theorem. Around a century ago it was realized that an extension to this type of dynamics can be obtained in the form of ``anomalous" diffusion where distributions are allowed to have heavy, power-law tails. Due to a unique feature of its momentum-dependant dissipative friction force, laser-cooled atomic ensembles can be used as a test bed for such dynamics. The interplay between laser cooling and anomalous dynamics bears deep, predictive implications for fundamental concepts in both equilibrium and non-equilibrium statistical physics. The high degree of control available in cold-atom experiments allows for tuning of the parameters of the friction force, revealing transitions in the dynamical properties of the system. Rare events, in both the momentum and spatial distributions, are described by non-normalized states using tools adapted from infinite ergodic theory. This leads to new experimental and theoretical results, illuminating the various features of the system.

\end{abstract}

\maketitle
{
  \hypersetup{linkcolor=black}
  \tableofcontents{}
}

\section{Introduction}
\label{Sec:Introduction}

Diffusive processes such as Brownian motion are ubiquitous in nature. In these, the position coordinates of an ensemble of non-interacting particles -- all starting from a common origin -- are normally distributed. This is a manifestation of the central limit theorem as the basic motion is composed of many random, uncorrelated steps. Mathematically, Brownian motion can be described either using a stochastic differential equation~\cite{gardiner1985handbook,VanKampen} that models the trajectory of a single particle, or with the diffusion equation relating to the description of the entire ensemble. The asymptotic time dependence of the mean squared displacement of such a motion is characteristically given by $\langle x^2 \rangle \sim t$, where \braket{...} denotes an average over the ensemble.

It is now well established~\cite{bouchaud1990anomalous,metzler2000random,Sokolov2005} that this is merely a particular case of a much richer set of phenomena where the dynamics differs in general from the ``normal" behavior, and $\langle x^2 \rangle \sim t^\beta$ where the exponent $\beta$ is not necessarily unity. Studying the mathematical properties of an extension to the standard central limit theorem, L\'evy~\cite{levy1937theorie} considered the problem of the summation of a large number of independent and identically distributed random variables whose variance diverges (Sec.~\ref{Section:IntroLevy}). This arises when the distribution of the random variables is heavy-tailed and decays as a sufficiently small power law. The resulting family of distributions describing the sum are called stable distributions, giving rise to the normal Gaussian as a special -- albeit clearly important -- case. Early physical manifestations of these include canonical work on spectroscopy~\cite{holtsmark1919verbreiterung} and the study of the distribution of gravitational forces acting on a tracer~\cite{chandrasekhar1943stochastic}. These statistical laws are now widely-used tools in many fields such as econophysics~\cite{mantegna1999}, soft-matter and biophysics~\cite{Shusterman2004,song2018neuronal}, dynamics of blinking quantum dots~\cite{stefani2009beyond} and hydrodynamics~\cite{kelly2019fractional}. 

One simple and elegant model allowing access to the full richness of the anomalous diffusion parameter space is a generalization of the celebrated L\'evy flight~\cite{mandelbrot1982fractal,shlesinger1995levy}. In this random-walk model, a walker ``jumps" a certain distance and then dwells at the arrived location for a certain time. Jump distances and dwell times are random variables, both drawn from distributions which can, in general, be heavy-tailed or have diverging moments~\cite{Scher1975}. Intuitively it can be understood that if the dwell time distribution has a heavy tail the dynamics can become ``sub"-diffusive ($\beta < 1$), whereas if the moments of the jump distances are allowed to diverge the dynamics may become ``super"-diffusive ($\beta > 1$). An important issue when dealing with such power-law distributions is that jumps can occur on many, widely different, scales. Since in physical processes long jumps are expected to take longer than short ones, and since a truly diverging jump length can never realistically be obtained, a modification to the original L\'evy approach was put forth. In the simplest version of this ``L\'evy walk" model~\cite{Shlesinger1987,Zaburdaev2015Review}, a fixed finite speed is prescribed, ensuring that long jumps take longer than short ones. More generally, a power-law correlation between the distance covered in a jump and the time duration of the jump is considered~\cite{Shlesinger1987}. Other generalizations of the L\'evy walk have also been proposed and analyzed as models of anomalous diffusion~\cite{Albers2018,Bothe2019,Vezzani2020}.

While much is known about the stochastic foundation of the L\'evy walk and its applications, a physical model that allows control over the transition between the phases of the dynamics could pave the way for a deeper understanding. As originally described in~\cite{Marksteiner1996}, laser-cooled atoms -- in certain parts of parameter space -- are ideal for this purpose. This realization gave rise to a series of questions regarding fundamental issues in statistical physics such as the non-Maxwellian nature of the velocity distribution, rare events in heavy-tailed processes, calculation of anomalous transport constants, and the ergodic properties of these non-equilibrium processes~\cite{Bardou2002,Lutz2013}. As a concrete example of such a fundamental concept, consider the two-time velocity autocorrelation function $\braket{v(t_2)v(t_1)}$. For Brownian motion and a vast number of other transport systems, this quantity is stationary and depends only on the time difference $\vert t_2-t_1\vert$. This has many consequences, for example for the calculation of the diffusion constant using the famed Einstein-Green-Kubo approach and also for the ergodic properties of the system. Under certain conditions, however, this stationarity property does not hold for laser-cooled atoms, and instead the correlations exhibit scale-invariance where the ratio of the two times becomes important. This leads to new ideas on transport and ergodicity.

Laser cooling is a well-established experimental technique for obtaining extremely low temperatures of atomic ensembles\footnote{The temperature of standard gasses is given by the width of the stationary, Gaussian momentum distribution. In the context of this work, however, momentum distributions typically deviate from Gaussianity and hence the definition is more subtle, and is discussed at length at the end of Sec.~\ref{Section:Momentum}.}. A momentum-dependent friction force is generated by external optical fields, reducing the atoms' momentum towards zero. The canonical example is that of Doppler cooling~\cite{Hansch1975,Phillips1998,Wineland1975,Wineland1978} where pairs of counter-propagating beams, detuned from the relevant two-level atomic resonance, selectively reduce the momentum of fast-moving atoms. The associated minimal Doppler temperature $T_D$ - brought about by the balance between the friction and random emissions of photons jolting the atoms - is given by $k_B T_D =\hbar \Gamma/2$, where $\Gamma$ is the natural linewidth of the excited state, $k_B$ is the Boltzmann constant and $\hbar$ is the reduced Planck constant. This was originally thought to be a fundamental limit for the ability to laser cool atomic systems and it was, therefore, a great surprise when temperatures below the Doppler limit were obtained experimentally~\cite{Chu1998Nobel,CCT1998Nobel} via a process later coined ``Sisyphus cooling" (Sec.~\ref{Section:IntroMechanism}). The key to explaining this discovery was the multilevel, degenerate nature of realistic atoms. The random momentum recoils $\hbar k$ due to scattered photons give rise to a second lower bound for the temperature, the recoil limit $k_B T_R/2 = \hbar^2 k^2 /2M = E_R$, where $E_R$ is the recoil energy, $M$ is the atomic mass and $k$ is the wavenumber. Temperatures even lower than the recoil limit have been achieved through other techniques such as sub-recoil laser cooling~\cite{Aspect1988,Bardou1994}, Raman and Raman-sideband cooling~\cite{Kasevich1992,Vuletic1998} and evaporative cooling~\cite{Anderson1995}, relying on different physical mechanisms.

For Sisyphus cooling, the particular dependence of the friction force on the atomic momentum leads to anomalous kinetics. The phase-space trajectory of an atom in the semiclassical approximation of the laser cooling mechanism has been analyzed theoretically with tools from quantum optics and the theory of stochastic processes~\cite{Dalibard1989,Marksteiner1996}. 

While anomalous dynamics can be found in many systems within the context of ultracold atomic physics~\cite{Kindermann2016Nature,Dechant2019,Niedenzu2011,Zheng2018,Meir2016}, in this Colloquium we focus on the recent developments in both theoretical and experimental understanding of the anomalous dynamics of atoms undergoing ``Sisyphus cooling". The model presented departs from the standard descriptions of L\'evy walks, which are usually postulated ad-hoc, and lays bare the mechanism behind the non-normal kinetics. From an experimental perspective, it allows unique control of the basic phenomenon enabling the transition between different phases of the dynamics. This is in stark contrast to most super-diffusive systems, say in the atmosphere or in the context of cell biology, where there is little experimental control over the basic features of the process. The main point of view is to present the modifications of the basic L\'evy walk model that emerge, and the consequences for basic concepts of equilibrium and non-equilibrium physics hand in hand with relevant experimental results. Among the topics discussed are the power-law distributions in momentum and position as well as their correlations (Sec.~\ref{Section:MomentumRealCorr}), the implications to such fundamental concepts in statistical physics as the Einstein-Green-Kubo relation, the breakdown of ergodicity and energy equipartition and the relation to infinite ergodic theory (Sec.~\ref{Section:FundamentalConcepts}). 


\section{L\'evy dynamics and Sisyphus cooling}
\label{Sec:Basics}

\subsection{L\'evy vs. Gauss central limit theorem, L\'evy flights and walks}
\label{Section:IntroLevy}

The process of Brownian motion was modelled by Einstein, Smoluchowski and others~\cite{Kubo1957,Hanggi2005,majumdar2007brownian}. It is very natural that the underlying random walk describing it is a Gaussian process as this is precisely what the central limit theorem predicts for a process where the total displacement at long times is a sum of many independent random displacements. Deviations from normal, Gaussian, Brownian motion can be parametrized by the random walk governed by L\'evy laws. 

Central limit theorems deal with the problem of summation of a large number $N$ of independent, identically-distributed random variables $\{\chi_i\}$. The sum $S=\sum_{i=1} ^N \chi_i/N^{1/\mu}$, scaled by some power $1/\mu$ of $N$ is considered. The probability density function (PDF) of $S$ is given by the inverse Fourier transform of its characteristic function,
\begin{align}
\begin{split}
  \left\langle \exp( i k S)\right\rangle &= \left\langle \exp\left(\frac{i k \chi_1}{N^{1/\mu}} \right)\right\rangle \cdots \left\langle \exp\left(\frac{i k \chi_N}{N^{1/\mu}} \right)\right\rangle\\ 
  &= \left\langle \exp\left(\frac{i k \chi}{N^{1/\mu}} \right)\right\rangle^N,
\end{split}\label{eq01}
\end{align}
where on the left the average is taken over the random variable $S$ and on the right it is taken with respect to the random variable $\chi$. The result is a direct outcome of the assumption that the random variables $\{\chi_i\}$ are independent and identically-distributed and so the expectation value factorizes. We will assume that the PDF of $\chi$ is symmetric so that its mean is zero. Two examples are the Gaussian PDF $P_G(\chi) =\exp(- \chi^2 / 2 \sigma^2)/\sqrt{2 \pi \sigma^2}$ with standard deviation $\sigma$, and a Lorentzian PDF $P_L(\chi)= [\pi (1+\chi^2)]^{-1}$. The first is an example of a distribution with finite moments, while the second has a power-law tail, and its variance diverges. The characteristic functions of the Gaussian and Lorentzian are $\tilde{P}_G(k) = \exp( - k^2 \sigma^2/2)$ and $\tilde{P}_L(k)=\exp(- |k|)$ respectively. Related to the divergence of the variance, the characteristic function of the Lorentzian exhibits non-analytical behavior at $k=0$, $P_L(k) \sim 1 - |k| $ and the second derivative with respect to $k$ at this point diverges. More generally, if $P(\chi)\sim |\chi|^{-(1 +\nu)}$ for large $|\chi|$ and $0<\nu<2$, then for small $k$ we have $\tilde{P}(k) \sim 1 - A |k|^\nu$, where $A$ is a scale parameter used as an input for the theory~\cite{bouchaud1990anomalous}. For $\nu<2$ the variance of the summand diverges. On the other hand, for any parent distribution of $\chi$ with a finite variance, we have $P(k) \sim 1 - \sigma^2 k^2/2$ where the leading term reflects the fact that $P(\chi)$ is normalized. Then, using Eq.~\ref{eq01}, two generic possibilities are found in the limit of $N\to \infty$. If the variance of $P(\chi)$ is finite, 
\begin{equation} 
  \left\langle \exp\left( i k S\right)\right\rangle \rightarrow \left(1 - \frac{\sigma^2 k^2}{2 N}\right)^N= \exp\left( - \sigma^2 k^2/2\right) \label{eq02} 
\end{equation} 
where we choose $\mu=2$ and use the definition of the exponential limit. This implies a diffusive scaling for the sum. Importantly, it also means that the PDF of the scaled sum $S$ is Gaussian for \emph{any} parent distribution with a finite variance $\sigma$. In contrast, if $P(\chi)$ exhibits power-law decay with $\nu<2$, then $\mu$ is chosen\footnote{Though a rigorous proof of the central limit theorem goes beyond the scope of this Colloquium, note that any other choice of $\nu\neq \mu$ would not give a meaningful limit when $N\to \infty$.} such that $\mu=\nu$, and
\begin{equation} 
  \left\langle \exp\left( i k S\right)\right\rangle \rightarrow \left(1 - \frac{A |k|^\nu}{N}\right)^N= \exp\left( - A |k|^\nu\right). \label{eq03} 
\end{equation} 
In this case super-diffusive scaling emerges, as the sum of the random variables grows like $N^{1/\nu}$. 

To summarize, the sum of random variables is scaled with $N^{1/\mu}$. Eq.~\ref{eq02} implies that if the variance is finite then $\mu=2$, while Eq.~\ref{eq03} together with the definition of the exponential function results in $\mu=\nu$ if $\nu<2$ (and the variance diverges). The inverse Fourier transform of Eq.~\ref{eq03} is the PDF of the sum $S$ and is also the well-known symmetric stable density or L\'evy density~\cite{bouchaud1990anomalous,Amir2020}-
\begin{equation} 
  L_{\nu,0} (S) = \frac{1}{2 \pi} \int\limits_{-\infty} ^{\infty} dk\exp( i k S - |k|^\nu), \label{Eq:SymLevy}
\end{equation} 
where the subscript $0$ indicates symmetric functions and the width scale $A$ is set to unity. In particular, the case $\nu=1$ is the Lorentzian and $\nu=2$ the Gaussian. 

\begin{figure} 
  \centering
    \begin{overpic}
      [width=\linewidth]{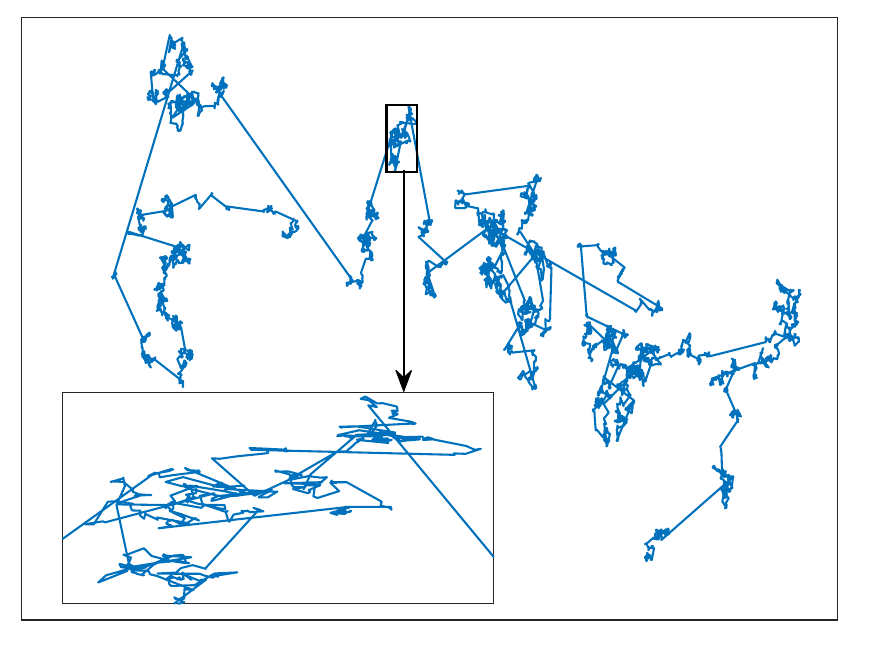}
    \end{overpic}
    \caption{L\'evy flights are random walk processes where the jump distances are obtained from heavy-tailed power-law distributions with infinite variance (Eq.~\ref{eq03}). Depicted is an example of a two-dimensional flight, with the $x$ and $y$ axes representing spatial dimensions and a L\'evy index $\nu = 1.4$. The direction of a given jump is rotated from that of the previous one by a random angle uniformly distributed between 0 and 2$\pi$. The fractal, self-similar nature of the dynamics is evident from the inset, similarly characterized by a small number of long flights and a large number of short ones.}
	{\label{fig:fig1}}
\end{figure}

There now emerge two forms of central limit theorems, the first is usually associated with Gauss\footnote{Though in modern physics this is called a Gaussian (1809), its roots precede Gauss and extend deep into the 18$^{\rm{th}}$ century with Bernoulli, de-Moivre and Laplace~\cite{stigler1986history}.} (Eq.~\ref{eq02}) and the second associated with L\'evy (Eq.~\ref{eq03}). When a system exhibits power-law statistics of the L\'evy type, the largest summand in the set $\{ \chi_i \}$ is of the order of the entire sum~\cite{chistyakov1964theorem,vezzani2019single}. The fact that the variance diverges implies that the underlying random walk is a fractal, also called a ``self-similar" object. This is easily visualised considering a 2d L\'evy flight, as depicted in Fig.~\ref{fig:fig1}. 

Specific examples of L\'evy stable densities emerged in physics, absent the general framework of mathematicians, in the context of the summation of a large number of forces or energies already more than a century ago. In his line-shape theory, Holtsmark~\cite{holtsmark1919verbreiterung} considered the distribution of the sum of many perturbations acting on an ion, and a similar problem was later discussed by Chandrasekhar~\cite{chandrasekhar1943stochastic} in an astrophysical context. The Holtsmark distribution is a particular, yet significant case of the L\'evy stable distribution, with $\nu=d/n$, where $d=3$ is the dimension of the system and $n=2$ is the power of the spatial dependence of the interaction potential, \ie gravitational or Coulomb. It applies when considering the projected force on one of the axes and when the bath particles are uniformly distributed in space. Landau, in his work on the energy transfer due to multiple scatterings in ionization~\cite{Landau1944}, constructed what would become known as the one-sided L\'evy stable distribution. Similar ideas and behaviors are found for single molecules in low temperature glasses~\cite{barkai2000levy,barkai2003levy}, and models of active dynamics~\cite{kanazawa2020loopy}. 

The notion of a L\'evy flight, where jump displacements are drawn from a common PDF with a power-law tail and summed, is problematic for a process describing spatial dynamics, as larger jump distances in space are expected to take longer. The power-law tail of the jump-size distribution means that the mean-squared displacement (MSD) of the L\'evy flight is infinite, which is unphysical as any real system will have a maximal speed of propagation. To remedy this, the concept of L\'evy \emph{walks} was introduced by~\cite{Shlesinger1987,Zaburdaev2015Review}. 

The simplest, one-dimensional, version of the L\'evy walk considers a particle starting at the origin at $t=0$. A random walk-duration $\tau$ is drawn from a heavy-tailed PDF $\psi(\tau)$. A random velocity is also drawn, for example $\pm V_0$ each with probability $1/2$. During this interval the motion is ballistic, reaching $x=\pm V_0\tau$. The process is then repeated (``renewed") and a new pair of walk duration and velocity is drawn. The position $x(t)$ of the particle is limited by the ballistic light cone $-V_0 t < x(t)< V_0 t$, and the moments of $x(t)$ never diverge for any finite $t$. If the distribution of walk durations does not have power-law tails, but is exponential, this model is essentially the Drude model for transport of electrons in metals and the diffusion is normal. In the L\'evy walk case, the walk duration PDF is heavy tailed, $\psi(\tau) \sim \tau^{-(1 + \eta)}$. Here, if $1<\eta<2$ the variance of the flight duration diverges, while if $0<\eta<1$ the mean flight duration also diverges. This yields three dynamical phases for the MSD~\cite{Zaburdaev2015Review}
\begin{equation} 
  \langle x^2 (t) \rangle \sim \left\{ \begin{array}{l l} t^2 & 0<\eta<1 \\ t^{3-\eta} & 1<\eta<2 \\ t & \eta>2. \end{array} \right. 
  \label{eq04} 
\end{equation} 
When $\eta<1$ the dynamics are ballistic, whereas for $1<\eta<2$ the spreading is super-diffusive. When the first two moments of the PDF of the jump duration are finite ($\eta>2$), normal diffusion is recovered. In the L\'evy walk, therefore, a natural cutoff is created by introducing the finite velocity, curing the unphysical divergence of the variance of displacement in the corresponding L\'evy flight. Another major difference between L\'evy walks and L\'evy flights is that in the latter the number of steps $N$ is fixed, like in any other random walk process, whereas in the L\'evy walk the number of renewals in the time interval $(0,t)$ is itself a random variable~\cite{godreche2001statistics}. 

There are many physical applications of L\'evy walks. For example, blinking quantum dots~\cite{Brokmann2003,MargolinPRL2005,stefani2009beyond} work in the ballistic phase $0<\eta<1$, where the ``effective velocity" is the intensity of emitted light which jumps between dark and bright states with power-law distributed sojourn times. The position of the random walker corresponds to the total number of photon counts which exhibits super-diffusive statistics, as seen in experiment~\cite{Margolin2006}. L\'evy walks appear also in the motion of bacterial colonies~\cite{Ariel2015} and in many other systems~\cite{Zaburdaev2015Review}. 

\subsection{The basics of Sisyphus cooling}
\label{Section:IntroMechanism}

The mechanism by which anomalous diffusion is manifested within the context of cold atomic ensembles in dissipative optical lattices\footnote{Not all atomic species trapped in near-resonant optical lattices would display such behavior. Bosonic Yb, for example, lacks the appropriate level structure and has no Sisyphus effect~\cite{Kostylev:14}} is related to the nonlinear nature of the momentum-dependent optical friction force $f(p)$ acting on the atoms~\cite{Dalibard1989,Castin1990,Agarwal1993,Marksteiner1996}. In the semi-classical approximation and for a given set of damping strength $\mathcal{A}$ and momentum capture range $p_c$, it takes the form~\cite{castin1991limits} 
\begin{equation} 
  f(p) = - \frac{\mathcal{A} p}{1+(p/p_c)^2}.
  \label{eq06} 
\end{equation} 
$\mathcal{A}$ and $p_c$ are functions of the experimental parameters of the system. A slow atom with $|p|/p_c\ll1$ will experience a drag-like force $f \sim - p$ similar to the Stokes friction acting on a Brownian particle in fluid, whereas a fast atom feels instead a weak force, $f \sim -1/p$. Intuitively, fast-moving atoms tend to remain fast, leading to large flights in space and in turn to anomalous L\'evy-type motion.

\begin{figure} 
  \centering
    \begin{overpic}
      [width=\linewidth]{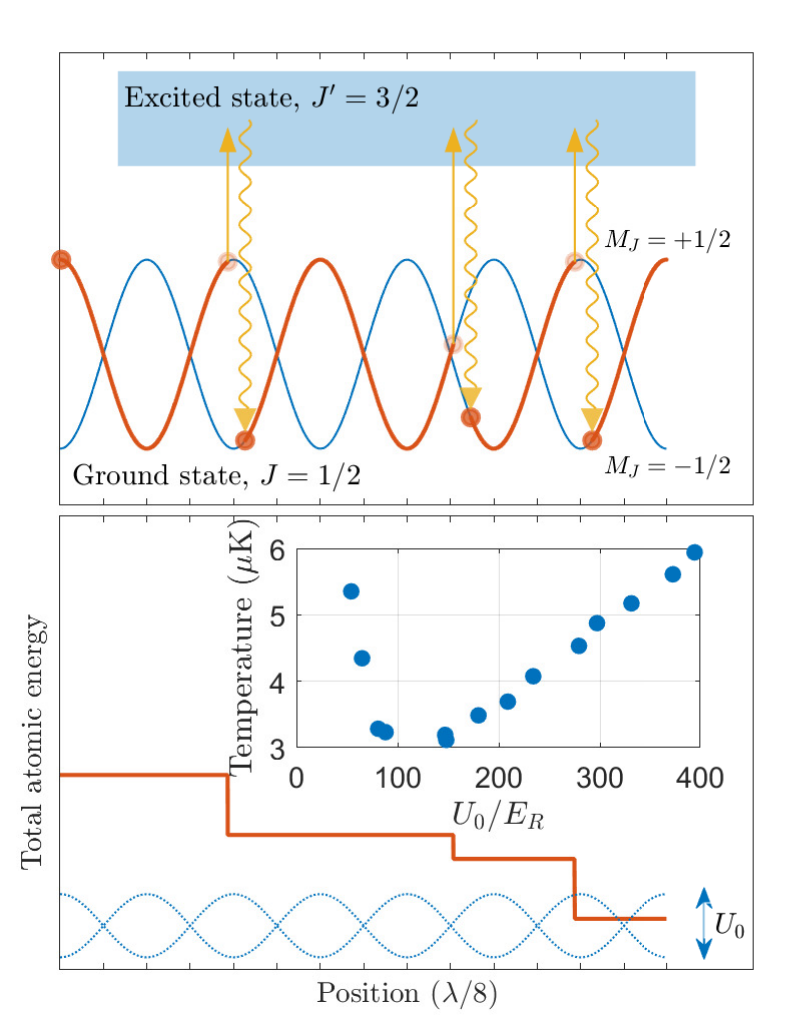}
    \end{overpic}
    \caption{Adapted from~\cite{castin1991limits}. L\'evy dynamics can arise in the motion of atoms undergoing \emph{Sisyphus cooling}. Top: Two counter-propagating, orthogonally-linearly-polarized laser beams generate a periodic spatial modulation of the polarization. The atomic ground state energy sub-levels $M_J=+1/2$ and $M_J=-1/2$ are then respectively perturbed by the standing light wave, so that the atom travels up and down hills and valleys of potential energy. When the laser frequency is tuned below the atomic resonance frequency, state-changing photon scattering is more probable around the top of the potential hills where the spontaneously emitted photon has a higher frequency than the absorbed one, illustrated as wiggly and straight yellow lines, respectively. As a result, an atom - moving from left to right in the depicted example - loses energy on average due to photon scattering, so that cooling becomes efficient (bottom). The main control parameter is the modulation-depth of the lattice $U_0$. The inset shows a measurement of the temperature of Sisyphus-cooled $^{133}$Cs atoms as the lattice depth is varied~\cite{Jersblad2000}, demonstrating a sharp increase of temperature when $U_0 / E_R$ decreases.}
	{\label{fig:fig2}}
\end{figure}

To realize this friction~\cite{CCT1998Nobel,Chu1998Nobel,foot2005atomic}, consider the qualitative picture of an atom that has a lower energy level with angular momentum $J = 1/2$ and an upper energy level with $J' = 3/2$ that moves through a standing wave formed by two counter-propagating laser beams with orthogonal linear polarizations (Fig.~\ref{fig:fig2}, top). The resulting polarization depends on the relative phase of the two laser beams and varies periodically with position $x$, changing from linear at, say, $x=0$ to $\sigma^-$ at $x=\lambda/8$ to orthogonal linear at $x=2\lambda/8$ to $\sigma^+$ at $x=3\lambda/8$. $\lambda$, of the order of a few hundreds of nm, is the wavelength of the lattice lasers. This polarization lattice causes periodic modulation of the states in the lower level manifold due to Stark shifts, enhancing the probability of the downhill transitions when the atom is at the top of the potential (and hence the term ``Sisyphus" following Greek mythology). The result is a net cooling effect as the atom, on average, slows down. The energy drops in Fig.~\ref{fig:fig2} (bottom) correspond to the absorption-emission events shown on the top panel. They are depicted as uneven to reflect the fact that the process is probabilistic and does not necessarily occur exactly at the peak. 

Sisyphus cooling reaches a non-equilibrium steady-state, where the friction force that biases the system towards zero momentum is balanced with the fluctuations caused by spontaneous emission. For ``deep" lattices, where the potential modulation depth $U_0$ defined in Fig.~\ref{fig:fig2} is large, the temperature is proportional to, and of the order of $U_0$~\cite{CCT1998Nobel}. This can be intuitively understood given that once the energy of the atom is smaller than $U_0$ it cannot climb up the hill, and hence at this stage cooling is ineffective (Fig.~\ref{fig:fig2}, bottom). As $U_0$ is reduced, the temperature reaches a minimum at some $U_0^{{\rm min}}$ and then rises sharply upon further reduction of $U_0$ (``shallow" lattices). At this minimum, to which typical experimental systems are tuned, $\braket{E_k}\sim U_0^{{\rm min}}$. 

Sisyphus cooling was extensively studied experimentally for a variety of atomic species. Typically pairs of counter-propagation red-detuned laser beams are used to generate the dissipative lattice, although three or four beams at proper angles can also be used~\cite{Kerman2000}, depending on the dimensionality of the problem. The lattice depth is controlled via the power and detuning of the cooling lasers, and magnetic fields must be kept below a few tens of mG to suppress shifts of the Zeeman levels that might hinder the Sisyphus effect. Finally, additional lasers are needed to ``repump" the atoms into the cycling transition levels but their direct effect on the atomic motion is usually negligible. In the deep-lattice regime, the temperature can be determined by measuring the standard-deviation of the velocity distribution, a task mostly performed using the time-of-flight technique, where a cloud of atoms is released from all confining fields and is allowed to freely expand. This generates a mapping of the atomic velocity distribution onto its spatial distribution after the expansion (assuming sufficient time elapses for the cloud to expand to a size much larger than its original size). The density profile of the expanded atoms, now describing their initial velocity distribution, is then imaged using fluorescence or absorption imaging. Additional techniques to measure atomic velocity distributions are based on velocity selective Raman transitions~\cite{Moler1992}, on spatial correlation functions~\cite{PhysRevLett.79.3146} or on measuring the survival probability after a sudden decrease of the trap depth [applicable for single atoms ~\cite{PhysRevA.67.033403,PhysRevA.78.033425}].

The inset of Fig.~\ref{fig:fig2} shows the result of such a time of flight temperature measurement after three-dimensional Sisyphus cooling of $^{133}$Cs atoms~\cite{Jersblad2000} as a function of the optical lattice depth $U_0$. The results show a minimal temperature at $U_0^{{\rm min}} / E_R \approx 100$. The distinction between shallow ($U_0 < U_{0} ^{{\rm min}}$) vs. deep ($U_0 > U_{0} ^{{\rm min}}$) lattices, that will be shown to mark the transition between different regimes in the dynamics, then follows naturally.

\section{Anomalous statistics and infinite densities}
\label{Section:MomentumRealCorr}

\subsection{Momentum space}
\label{Section:Momentum}

We now turn to a quantitative description of the anomalous dynamics of the momentum distribution for 1d motion in the dissipative lattice, after averaging over the lattice period and using the semiclassical approximation. Generally this is valid in the shallow lattice regime and ignores trapping in the wells of the optical lattice. More specifically, it requires that \emph{(i)} The laser is weak, meaning that the saturation parameter $s_0\ll 1$ ($s_0$ is a measure of the occupation of the excited state proportional to the laser intensity~\cite{foot2005atomic}). \emph{(ii)} The atomic kinetic energy is large compared to the lattice depth such that all positions along the lattice are considered equiprobable allowing spatial averaging. \emph{(iii)} The atomic momentum change $\Delta p\gg\h k$, where $k$ is the wavenumber of the laser field~\cite{Lutz2003}. The validity of the semiclassical approximation has been tested using full quantum Monte-Carlo wave function simulations~\cite{castin1991limits,Marksteiner1996} and it has been shown to quantitatively hold under the conditions described above, assuming an ideal atomic level structure of $J=1/2\to J'=3/2$. For other level structures the results show a qualitative agreement.

This semiclassical treatment is a useful approximation since the long jumps giving rise to the L\'evy diffusion are not influenced by the trapping potential. The procedure of the derivation of the Fokker-Planck equation involves a spatial averaging over the wells, namely it is assumed that spatial modulation of the optical lattice has marginal effect on the anomalous statistics~\cite{Marksteiner1996,castin1991limits}. In the opposite limit of deep lattices, the spatial structure of the lattice and the energy surfaces play an important role and cannot simply be averaged out.

It has been shown~\cite{castin1991limits,hodapp1995three,Lutz2004} that the dynamics of the momentum PDF $W(p,t)$ is governed by the following Fokker-Planck equation\footnote{A first derivation of the Fokker-planck equation for atoms in a field of laser radiation pressure can be found in~\cite{Letokhov1981}. This equation is derived in the semiclassical limit, using an expansion of the master equation in terms of the recoil velocity.},
\begin{equation} 
  \frac{\partial W }{ \partial t} = - \frac{\partial }{ \partial p} \left[ f(p) W \right] + \frac{\partial }{ \partial p} \left[ D(p) \frac{\partial W }{ \partial p} \right], 
  \label{eqLutz} 
\end{equation} 
valid for shallow lattices.  The diffusive term $D(p) = D_0 + D_1 /[1 + (p/p_c)^2]$ describes stochastic fluctuations of the momentum where $D_0$ and $D_1$ are functions of the experimental parameters~\cite{Castin1990,Marksteiner1996}. Unlike the friction force that vanishes in the limit of large momentum, the momentum diffusion becomes $p$-independent in this limit, and $D(p) \to D_0$. The simple relation between friction and dissipation, in the spirit of the Einstein relation, is hence invalid.

The steady-state solution of Eq.~\ref{eqLutz}, $W(p)$, was derived in~\cite{Lutz2003}. Using the force given by Eq.~\ref{eq06} and the diffusive term discussed above, it reads 
\begin{equation} 
  W(p) \sim \left[ 1 + \frac{ D_0}{D_0 + D_1} \left(\frac{p}{p_{c}}\right)^2 \right]^{-\widetilde{U}_0/2}.
  \label{eq05} 
\end{equation} 
The width of the momentum distribution is determined by $p_c$, and it has a power-law tail with an exponent expressed in terms of $\widetilde{U}_0$ as $W(p) \sim |p|^{-\widetilde{U}_0}$. Emerging from the Fokker-Planck equation as an important dimensionless parameter in the system, $\widetilde{U}_0$ is defined in terms of the experimental parameters as 
\begin{equation} 
    \widetilde{U}_0 = \mathcal{A}\frac{p_c^2}{D_0}= \frac{1}{C}\frac{M\delta s_0}{\hbar k^2}= \frac{1}{C}\frac{U_0}{E_R} . \label{eq:defU0} 
\end{equation} 
Here $\delta$ is the detuning of the laser from the atomic transition frequency. Different values are cited in the literature for the proportionality constant $C$, reflecting the complexity of experimental atomic systems beyond the simplified models~\cite{Castin1990,Marksteiner1996}. The exact numerical value does not, however, have a profound effect on the results presented here, and therefore it is reasonable to treat $C$ as a dimensionless fitting parameter. Even though both the detuning of the laser from the atomic resonance and its intensity affect the potential depth as well as the photon scattering rate, the effective temperature and indeed the anomalous dynamics only depend on the single parameter $\widetilde{U}_0$.

\begin{figure} 
  \centering
    \begin{overpic}
      [width=\linewidth]{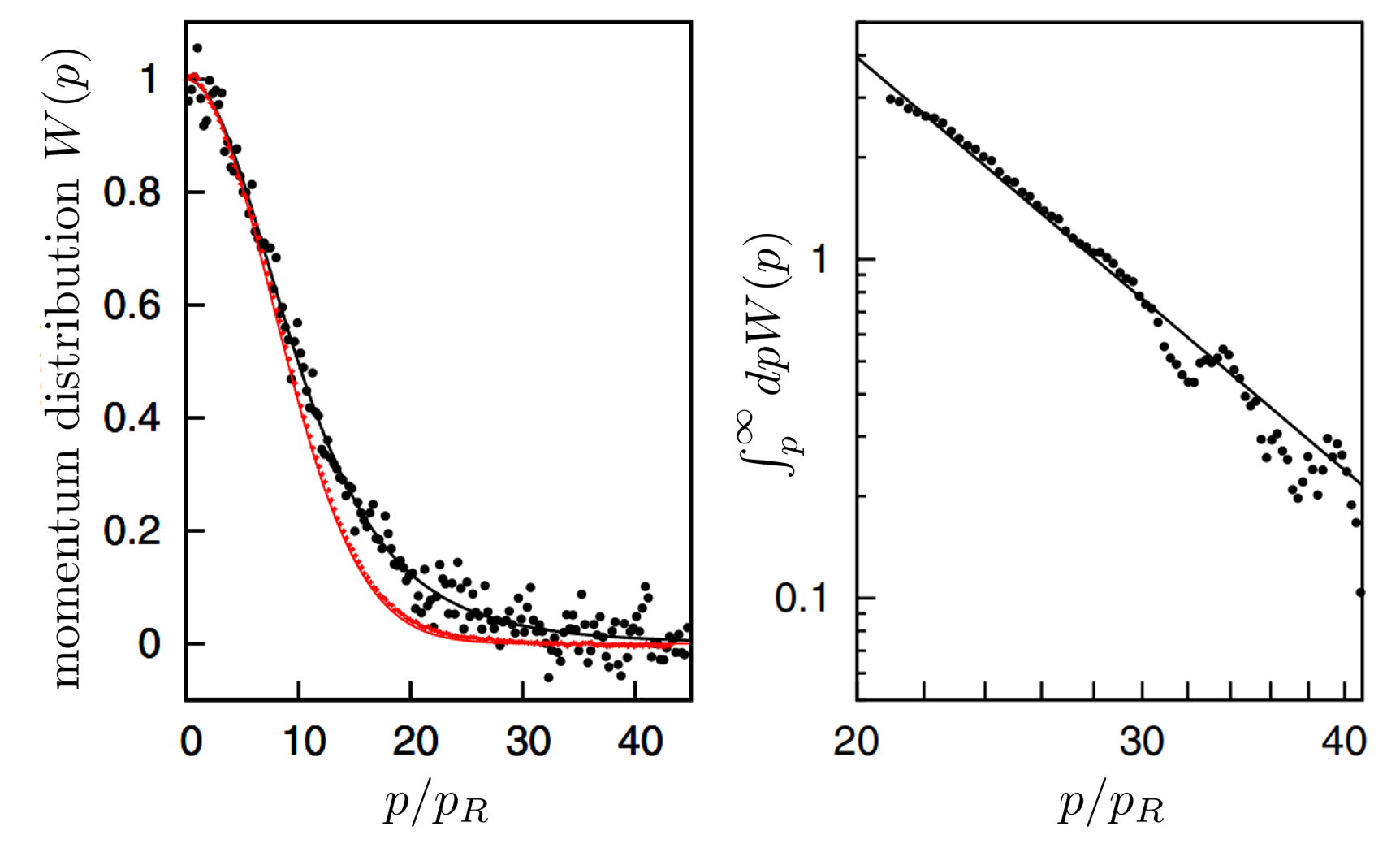}
    \end{overpic}
    \caption{Adapted from~\cite{Douglas2006}. Analysis of the atomic dynamics under the unique Sisyphus friction (Eq.~\ref{eq06}) predicts a non-Gaussian steady-state in momentum (Eq.~\ref{eq05}). The left panel shows the results of a time-of-flight measurement of the momentum distribution $W(p)$ of an ensemble of $^{133}$Cs atoms as a function of their momentum, rescaled by the recoil momentum $p_R=\sqrt{2ME_R}$. The plot compares a ``deep" lattice (red markers, almost coinciding with red line) and a ``shallow" lattice (black points). Solid lines represent fits to Eq.~\ref{eq05}, giving a Gaussian distribution for the deep lattice (red) and a power-law for the shallow lattice (black). (right) Integrated data from the left panel describing the probability for the momentum to be larger than some value $p$. Plotted on a log-log scale it reveals the power-law nature of the distribution.}
	{\label{fig:fig3}}
\end{figure}

The second moment of the steady-state momentum \braket{p^2}, usually considered as a measure of the temperature, diverges when $\widetilde{U}_0 < 3$. Moreover, when $\widetilde{U}_0 < 1$ the solution itself is no longer normalizable and there is no steady-state at all. The first experimental verification of Eq.~\ref{eq05} was presented in~\cite{Douglas2006}, where an ensemble of $^{133}$Cs atoms was exposed to a 3d Sisyphus lattice of variable depth\footnote{Although Eq.~\ref{eq05} is derived for 1d, in isotropic cases it can also apply to higher dimensions~\cite{Lutz2003}}. The atomic momentum distribution, measured by time-of-flight, is presented in Fig.~\ref{fig:fig3}. The results, representing an average over 200 separate images where extra care was taken to balance the radiation-pressure force from the counter-propagating beam pairs, led the authors to conclude that they represent a statistically significant indication of the power-law tails of the momentum distribution. This can be seen in the right panel of Fig.~\ref{fig:fig3}, where the probability for an atom to have momentum larger than some value $p$ is shown to fit a power-law over two decades. The non-Gaussianity of the momentum distribution of atoms in the deep lattice regime was studied experimentally in~\cite{Jersblad2004} using $^{133}$Cs. They fitted time-of-flight data to several test functions, including the power-law type distribution of Eq.~\ref{eq05} and a double Gaussian, and concluded that a double Gaussian provides a better fit to the experimental data. 

In analyzing the dynamics described in Eq.~\ref{eqLutz}, it is convenient to also consider the corresponding Langevin equation representing the phase space trajectory of a single particle. There is a standard procedure for obtaining the corresponding Langevin equation from a given Fokker-Planck equation~\cite{VanKampen}, wherein the drift term corresponds to an external force and the diffusion term arises from the external noise. The analysis of trajectories is extremely useful for a myriad of reasons, ranging from ease of simulations that allow insight into the nature of individual trajectories to more subtle properties of the system such as the analysis of time averages to tackle issues like ergodicity (discussed at length in Sec.~\ref{Section:FundamentalConcepts}). For simplicity, we set $D_1=0$, since it modifies neither the asymptotic $|p|\to \infty$ behavior of the diffusive term nor the cooling force, and therefore does not affect the main conclusions. Transforming to dimensionless time $t \to \mathcal{A} t$, momentum $p \to p/p_c$ and position $x \to x M \mathcal{A}/p_c$, the Langevin equation, with $\xi(t)$ being white Gaussian noise with zero mean and second moment $\braket{\xi(t)\xi(t')}=\delta(t-t')$, reads~\cite{Barkai2014} 
\begin{equation} 
  \frac{d p}{d t} = f(p) + \sqrt{\frac{ 2}{ \widetilde{U}_0} }\, \xi(t) ; \ \ \frac{d x}{d t} = p
  \label{eq:Langevin}
\end{equation} 
where the dimensionless form of the force of Eq.~\ref{eq06} is $f(p) = -p/(1+p^2)$. 

The non-normalizability of the steady state for $\widetilde{U}_0<1$ points to a dynamical transition. The time it takes a particle with momentum $p>0$ to cross $p=0$ is random and the PDF of these times is described by a power-law, $\psi(\tau) \sim \tau^{ - (1+\eta)}$, with $0<\eta\le 1$ for $\widetilde{U}_0\le1$~\cite{Barkai2014,Marksteiner1996}. The non-normalised steady-state is thus related to the divergence of the mean return time of the momentum. Even though the energy cannot realistically diverge, measurements of the ensemble-averaged kinetic energy $\langle p^2 \rangle/2 M$ do exhibit a sharp transition at a certain $\widetilde{U}_0$ below which the energy increases dramatically (Fig.~\ref{fig:fig2}, inset). The divergence of the return time, and the vanishing of the normalization of the steady state, are generic themes that often appear together in other systems~\cite{Aghion2019}.

A theoretical challenge now arises. Assuming the Fokker-Planck equation is valid, is the associated steady-state $W(p)$ a valid description over times which are long but finite? It turns out that the steady-state by itself does not yield a complete description of the momentum distribution. In particular, all moments of the momentum distribution as specified by the time-dependent Fokker-Planck equation are finite for any finite measurement time, in contrast to the steady-state prediction made based on Eq.~\ref{eq05}. To obtain these moments, including the second moment (which as noted gives the ensemble-averaged kinetic energy) for $\widetilde{U}_0<3$, a new tool is needed, called the \emph{infinite covariant density}. 

\begin{figure} 
  \centering
    \begin{overpic}
      [width=\linewidth]{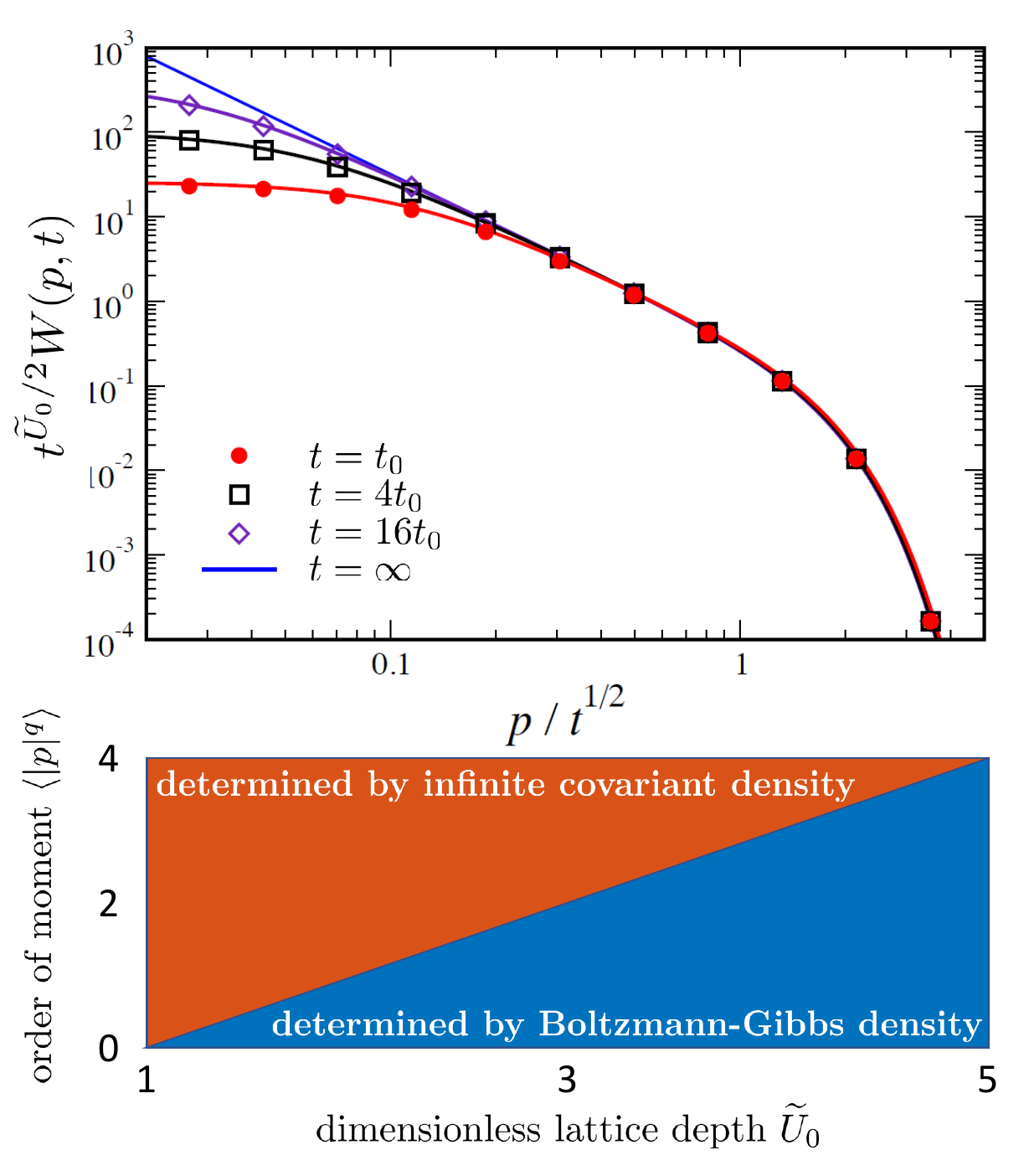}
    \end{overpic}
    \caption{(top) Adapted from~\cite{Kessler2010}. Temporal dynamics of the momentum distribution. The rescaled density of momentum versus $p/t^{1/2}$ obtained from Langevin simulations (Eq.~\ref{eq:Langevin}) with $\widetilde{U}_0=2$ for rescaled times $t = t_0 \equiv 76.3, t = 4t_0$ and $t=16t_0$, emphasizing the large momentum of the atom where the scaling is effectively diffusive as the friction is small. The theoretical curve is the \emph{infinite covariant density} of Eq.~\ref{eq:Wsolf}, a valid description in the long time limit. It exhibits a Gaussian-like cutoff which renders the moments finite, unlike those of the formal steady-state which diverge. (bottom) Adapted from~\cite{Holz2015}. The $q^{\rm{th}}$ moment of the momentum as a function of $\widetilde{U}_0$, showing the complementarity of the two densities in calculating moments of the observables. The steady state is (Eq.~\ref{eq07}) termed ``Boltzmann-Gibbs" since it describes an effective motion in a logarithmic potential (see text).}
	{\label{fig:fig4}}
\end{figure}

\paragraph*{Infinite covariant density.} 
It has been shown that the time-independent solution is not an adequate description of the system, as it gives rise to infinite moments. Of course, the full time-dependent solution gives physically meaningful answers for all finite times, however its exact form is not analytically attainable. A useful approximate result can be obtained if the problem is broken in two, focusing separately on the regimes $p\ll \sqrt{t}$ and $p \sim \sqrt{t}$. For the former case, the equilibrium answer is a good approximation. For the latter case, however, a new type of approximation is obtained, which amounts to a non-normalizable density. Such a seemingly paradoxical mathematical object has been given the name ``infinite density", and will be shown to be a crucial tool for correctly generating the second and higher moments of the momentum distribution.

The dimensionless form of the steady-state momentum distribution corresponding to Eq.~\ref{eq05} is 
\begin{equation} 
  W(p) = {\cal N} (1 + p^2)^{ - \widetilde{U}_0/2} \ \ \mbox{for} \ \ \widetilde{U}_0 > 1 
  \label{eq07} 
\end{equation} 
with ${\cal N} = \Gamma\left(\frac{\widetilde{U}_0}{2}\right) / \left[ \sqrt{\pi} \Gamma\left(\frac{\widetilde{U}_0 -1}{2}\right)\right]$. This describes the steady state of the process defined in Eq.~\ref{eq:Langevin} with $\Gamma(x)$ being the Gamma function. As mentioned above, for $\widetilde{U}_0\le 1$ the solution is no longer normalizable. For $\widetilde{U}_0>1$, the steady-state second moment is 
\begin{equation} 
  \langle p^2 \rangle = \left\{ \begin{array}{c c} \frac{1}{ \widetilde{U}_0 - 3} & \ \ \mbox{for} \ \ \widetilde{U}_0>3 \\ \infty & \ \ \mbox{for} \ \ \widetilde{U}_0 < 3. \end{array} \right.
  \label{eq:InfiniteEnergy}
\end{equation} 
The divergence of the steady-state kinetic energy $\langle p^2 \rangle$ as $\widetilde{U}_0$ approaches its critical value from above is the direct result of the power-law tail of the momentum distribution, which in turn is due to the weak friction force at large momentum $f(p) \sim - 1/p$. The key insight is that for these types of systems the difference between the steady-state distribution and the distribution at a large but finite time is non-negligible, in contradiction to what happens for the standard $f(p)\sim -p$ friction force. While the bulk of the distribution is given correctly by the steady-state solution, the power-law tail of the time-dependent $W(p,t)$ does not extend to infinite momentum. Just as in the case of free diffusion in momentum space where the momentum very rarely exceeds $\sim\sqrt{t}$, so too here the power-law tail is cut off at $|p| \sim \sqrt{t}$. To see this, a careful analysis of the dimensionless Fokker-Planck equation 

\begin{equation} 
  \frac{ \partial W}{\partial t} =\left( \frac{1}{\widetilde{U}_0} \frac{\partial^2}{\partial p^2 } + \frac{\partial}{\partial p} \frac{ p}{1 + p^2}\right) W 
\end{equation} 
was presented in~\cite{Kessler2010,dechant2011solution,Levine_2005} employing the scaling ansatz
\begin{equation} 
  W(p,t) \sim t^q h( p /\sqrt{t}) 
  \label{eq:Wscaling}
\end{equation} 
which holds for large momentum and long time. Using the diffusive scaling variable $z=p/\sqrt{t}$, the following equation is found-
\begin{equation} 
  \frac{1}{\widetilde{U}_0} \frac{d^2 h}{d z^2} +   \left( \frac{1}{z} + \frac{z}{2} \right) \frac{d h}{d z} - \left(   q + \frac{1}{z^2} \right) h = 0. 
\end{equation} 
Matching this solution to the steady-state solution which holds for $p\ll \sqrt{t}$ gives $q= - \widetilde{U}_0 / 2$, and 
\begin{equation} 
  h(z) = \frac{{\cal N} z^{-\widetilde{U}_0}}{\Gamma\left( \frac{\widetilde{U}_0 + 1}{2} \right) } \Gamma\left( \frac{ 1 + \widetilde{U}_0}{2} , \frac{\widetilde{U}_0 z^2}{4} \right) 
  \label{eq:Wsolf}
\end{equation} 
where $\Gamma(a,x)$ is the incomplete Gamma function. In the small and large $z$ limits,
\begin{equation} 
  h(z) \sim \left\{ \begin{array}{c c} {\cal N} z^{-\widetilde{U}_0} \ \ & z\ll 2/\sqrt{\widetilde{U}_0} \\ \frac{{\cal N} ( 4/\widetilde{U}_0)^{(1-\widetilde{U}_0)/2}}{\Gamma[(\widetilde{U}_0 + 1)/2]} z^{-1} e^{ - \widetilde{U}_0 z^2 /4 } \ \ & z\gg 2/\sqrt{\widetilde{U}_0} . \end{array} \right. 
  \label{eq:Wsolfasym}
\end{equation} 
The Gaussian factor found for large $z$ stems from the diffusion in momentum space, as the force becomes negligible for large momentum $p\gg \sqrt{t}$. The small-$z$ behavior is a power-law that matches the large-$p$ behavior of the steady state of Eq.~\ref{eq07}.  

The solution $h(z)$ is non-normalizable, since $h(z)\sim z^{-\widetilde{U}_0}$ for small $z$. This type of solution is called an infinite covariant density. It is \emph{covariant} in the sense that $z =p/\sqrt{t}$, hence $p$ must be scaled with the square root of time [namely $h(z)$ remains unchanged as both $\sqrt{t}$ and $p$ are modified, keeping their ratio fixed], while the normalized steady-state, Eq.~\ref{eq05}, is clearly time-\emph{invariant}. More importantly, the term infinite refers to the non-normalizablibility of the solution. Thus $h(z)$, while remaining positive, is certainly not a probability density. Its statistical meaning can be understood following the mathematical literature on infinite ergodic theory~\cite{aaronson1997introduction}. In the long time, large momentum limit, with $p/\sqrt{t}$ fixed-
\begin{equation} 
  \lim_{\substack{p,t \to \infty\\p/\sqrt{t}\  \rm{fixed}}} t^{\widetilde{U}_0/2} W(p,t) = h(p/ \sqrt{t}).
  \label{eq:U0lt1}
\end{equation} 
The right-hand side is not normalized since the perfectly normalized PDF $W(p,t)$ is multiplied by $ t^{\widetilde{U}_0/2}$ which diverges as $t \to \infty$. The function $h(z)$ can be used to compute the moments of the process $p(t)$, namely those that diverge with respect to the integration over the steady-state. For example, the second moment is 
\begin{equation} 
  \langle p^2 \rangle = \left\{ \begin{array}{c c} \frac{ 1}{\widetilde{U}_0 -3} \ \ \ & \widetilde{U}_0 > 3 \\ \frac{ 14 {\cal N}}{2^{\widetilde{U}_0} \Gamma\left( \frac{ \widetilde{U}_0 +1}{2} \right)} \frac{ 1}{2 - \widetilde{U}_0} (t/\widetilde{U}_0)^{ (3-\widetilde{U}_0)/2} \ \ & 1< \widetilde{U}_0 < 3\\  \frac{2(1-\widetilde{U}_0)}{\widetilde{U}_0}t \ \ \ \ & \widetilde{U}_0 <1 . \end{array} \right. 
  \label{eqppp} 
\end{equation} 
For $\widetilde{U}_0>3$ the kinetic energy is time-independent, determined by the steady-state solution, and blows up as $\widetilde{U}_0 \to 3$. For the intermediate range, $1<\widetilde{U}_0<3$, the behavior is sub-diffusive and the infinite density determines the kinetic energy. Even though for $\widetilde{U}_0<1$ the system is actually heating linearly with time, the infinite kinetic energy of Eq.~\ref{eq:InfiniteEnergy} is never obtained. At the critical value $\widetilde{U}_0=1$ the system transitions to normal diffusive scaling of the mean-squared momentum. Simulations and calculations taking account of the underlying lattice structure of the laser field were performed in~\cite{Holz2015}. These simulations solved the relevant set of stochastic differential equations for shallow lattices showing agreement with the infinite covariant density of Eq.~\ref{eq:Wsolf} and proving that the lattice structure does not affect the scaling properties of the system.

Infinite densities are important in many applications beyond Sisyphus cooling, in particular in chaos theory~\cite{akimoto2008generalized,korabel2009pesin}. They have been studied extensively in the context of infinite ergodic theory~\cite{aaronson1997introduction} where the behavior of time averages is important. For the system at hand, the steady-state solution and the infinite covariant density are complementary tools as both are long-time solutions of the problem. $W(p,t)$ converges towards the steady-state as long as $\widetilde{U}_0 > 1$. Similarly, plotted in the scaling form, the solution approaches the non-normalized infinite density (Fig.~\ref{fig:fig4}). For $1<\widetilde{U}_0 <3$, the steady-state solution predicts an infinite energy but a finite normalization and the infinite density gives a finite energy but infinite normalization. Hence both tools are required for a complete description of the dynamics [Fig.~\ref{fig:fig4} (bottom)].

Consider now the case of $\widetilde{U}_0<1$. The steady-state (Eq.~\ref{eq07}) is not normalizable, and now an infinite invariant density describes the momentum distribution in the inner region $p < \sqrt{t}$, and there exists a limit 
\begin{equation} 
  \lim_{t \to \infty} t^{ (1 - \widetilde{U}_0 )/2} W(p,t) \sim (1 + p^2)^{- \widetilde{U}_0 /2}. \label{eq20}
\end{equation} 
The expression on the right-hand side is a non-normalizable function since $\widetilde{U}_0<1$, however it still describes a long-time limit of the density~\cite{dechant2011solution}. The divergence emanates from the large-$p$ behavior. Since the right-hand side of Eq.~\ref{eq20} is time-independent, it is termed ``invariant" and not covariant as was the case of Eq.~\ref{eq:U0lt1}. It is not a coincidence that it is similar to the form found for the normalized steady-state (Eq.~\ref{eq07}). Here infinite ergodic theory comes into play, distinguishing between two types of observables: those that are integrable with respect to the non-normalized density (given by the time-independent right hand side of Eq.~\ref{eq20}) and those that are not\footnote{To clarify the concept of integrability consider a PDF $f(x)$ of a finite random variable $x$, and an observable $O(x)$. If the integral $\int_{-\infty} ^\infty O(x) f(x) d x $ is finite then the observable $O(x)$ is integrable. A similar concept is used even when the system is described by a non-normalised state.}.

This is similar to the kinetic energy observable considered for the case $\widetilde{U}_0>1$ which, depending on the value of $\widetilde{U}_0$, may be either integrable or non-integrable with respect to the infinite covariant density. The fluctuation behavior of the time-averages of observables was studied in~\cite{Aghion2019}. The mean of the time averages can be evaluated from the non-normalized state, at least for observables that are integrable with respect to the infinite density. This is somewhat similar to usual ergodic theory where the time averages are calculated with ensemble averages, however now instead of using a normalized distribution in the steady state, a non-normalizable function is used. Employing the Darling-Kac theorem~\cite{DarlingKac} and infinite ergodic theory it is found that certain observables, when time averaged, have a universal distribution which in turn is related to L\'evy statistics~\cite{aaronson1997introduction,Aghion2019}. The details of this, however, extend beyond the scope of this Colloquium. 

The regime $\widetilde{U}_0<1 $ is predicted to exhibit special features, and experiments in this region are technically challenging due to the fact that the system is actually being heated and atoms are easily lost from the trap. The issue of heating and loss of atoms can be partially mediated by use of elongated dipole ``tube" traps or blue-detuned optical box potentials~\cite{Sagi2012,Afek2017Correlations,Afek2020,navon2021quantum}, however so far there is no clear-cut experimental proof of Eq.~\ref{eq20}.

\paragraph*{The meaning of temperature.} When a gas is coupled to a heat bath, its equilibrium temperature is proportional to the variance of its momentum distribution which is Gaussian. How, then, does one define a temperature for the Sisyphus-cooled system? One approach is that there is no temperature at all since the system is in a non-equilibrium state. On a more practical level, though, a temperature-like quantifier can be defined which is the full width at half maximum (FWHM) of the momentum distribution, typically used in experiments. A third option is to use the mean of the kinetic energy, equivalent in equilibrium to the normal temperature. For shallow lattices, this observable should be obtained from the infinite invariant density and not from the steady state. Furthermore, in the shallow lattice regime the distribution can have a very narrow FWHM, but also a variance that increases with time. This means that one measure of the temperature (FWHM) can indicate that the system is cold while another indicates that it is hot.

\paragraph*{Diffusion in a logarithmic potential.}

The problem of momentum dynamics of Sisyphus-cooled ultracold atoms is related to the over-damped Langevin dynamics in a logarithmic potential~\cite{poland1966phase,bray2000random,PhysRevLett.98.070601,fogedby2007dynamics,bar2007loop,bar2008dynamics,Dechant2011,Hirschberg2011,Hirschberg2012,ray2020diffusion}. Intuitively, this connection stems from the fact that the friction force $f(p)\sim - 1/p$ at large $p$, and hence asymptotically the effective potential in momentum space is $V(p)= -\int dpf(p)  \sim \log(p)$. More explicitly, consider a Brownian particle in a potential $V(x)= V_0 \log (1 + x^2)$ in contact with a standard thermal heat bath with temperature $T$. According to the Boltzmann-Gibbs framework, the density in thermal equilibrium is $P_{{\rm eq}}= {\cal N} (1 + x^2)^{ - V_0/k_B T}$, bearing the same structure of the steady-state Eq.~\ref{eq05} with appropriate adjustments. The stochastic dynamics in a log potential is important for several problems, like DNA looping~\cite{Hanke2003,bar2007loop} and Manning condensation~\cite{Manning1969}.
 
\subsection{Position space}
\label{Section:Position}

\begin{figure} 
  \centering
    \begin{overpic}
      [width=\linewidth]{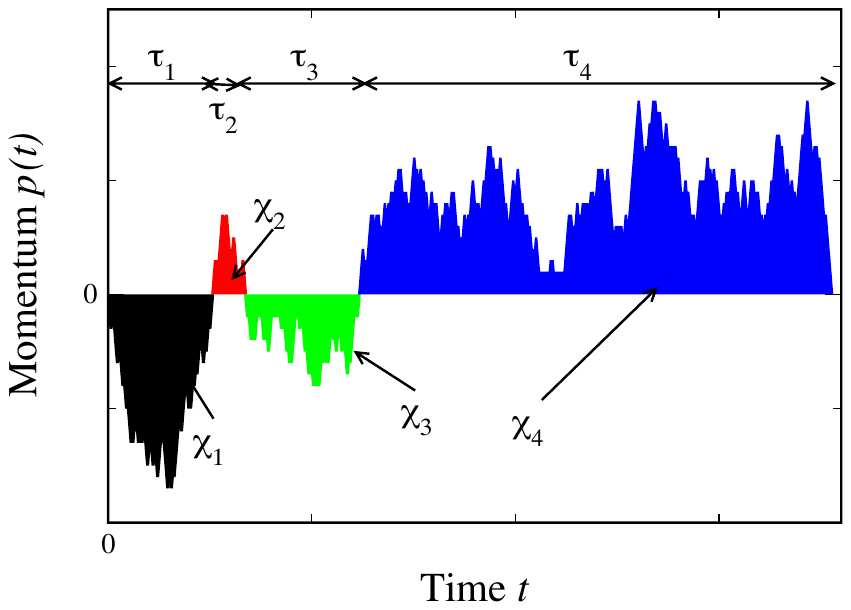}
    \end{overpic}
    \caption{Adapted from~\cite{Kessler2012}. The nature of the spatial diffusion is revealed by numerical simulation of momentum zero-crossing dynamics. The crossing of zero momentum defines durations $\tau_i$ and jump distances $\chi_i$, enabling identification of the connection between the continuous momentum trajectory of the atom and the L\'evy walk picture. The jump distances are the areas under the stochastic momentum curve which starts and ends at zero momentum and never crosses in between. Time periods between zero crossings are indicated by the black arrows on the top part of the plot.}
	{\label{fig:fig5}}
\end{figure}

An immediate consequence of the anomalous dynamics in momentum space is nontrivial dynamics in position space. In~\cite{Marksteiner1996}, the spatial diffusion of atoms was theoretically studied, revealing that below a critical depth of the optical lattice there exists a transition to a L\'evy-like motion. To show this, certain modifications of the basic L\'evy walk are needed. Consider a long-time momentum-space trajectory $p(t)$, crossing zero many times. Let $\tau$ be the random interval of time between two such successive crossings, and $\chi$ be the random displacement for a given such interval, schematically presented in Fig.~\ref{fig:fig5}. A process is generated with a set of random jump durations $\{\tau_i\}$ between zero crossings and corresponding displacements $\{ \chi_i\}$ given by $\chi_1 = \int_0^{\tau_1} dtp(t) $, $\chi_2=\int_{\tau_1}^{\tau_1 + \tau_2 }dt p(t)$, etc. For a particle starting at the origin at $t=0$ with $p=0$, the sum of all the displacements $\chi_i$ is the random position of the particle at time $t=\sum_i \tau_i$, denoted $x(t)$, and the corresponding PDF is $P(x,t)$. Since the Langevin process (Eq.~\ref{eq:Langevin}) is driven by white noise, defining zero crossings requires a more precise treatment~\cite{Kessler2012,Barkai2014,Majumdar2004,Majumdar2005}. We define the random time $\tau$ as the time it takes the particle starting with momentum $p=\epsilon$ to reach $p=0$ for the first time.

The Langevin process of Eq.~\ref{eq:Langevin} is Markovian, so once the particle crosses zero momentum, a given interval $\tau$ is independent of the previous one. This is nothing but a renewal process: $\tau_i$ is drawn repeatedly from $\psi(\tau)$, the crossing time PDF computed using the Langevin Eq.~\ref{eq:Langevin}, until the sum of the times exceeds the measurement time. The distribution of crossing times can be calculated using the tools of first-passage theory~\cite{redner2001guide,Miccich__2010,Martin2011}. The marginal PDFs of step durations and jump distances exhibit power-law behaviors~\cite{Marksteiner1996,Barkai2014} 
\begin{equation} 
  \psi(\tau) \sim \tau^{- (3/2) - \widetilde{U}_0/2}, \ \ \ q(\chi) \sim |\chi|^{-(4/3) - \widetilde{U}_0/3}. 
  \label{eqZoller} 
\end{equation} 
This scaling is the result of the weak friction found at large momentum. In the limit $\widetilde{U}_0=0$, $\psi(\tau)\sim \tau^{-3/2}$ which describes the distribution of first-passage times of a Brownian particle in 1d searching for a target at the origin~\cite{redner2001guide}. This extremely shallow lattice limit describes a pure diffusive process in momentum space with no friction. Long intervals are associated with large momenta and so the cooling friction does not destroy the existence of power-law decay at large $\tau$, though it does change the actual exponent. The mean of $\tau$ diverges when $\widetilde{U}_0\le 1$ and the mean first-passage time for the momentum, namely the time it takes the momentum to cross the origin, becomes infinite when the steady-state of the momentum distribution Eq.~\ref{eq07} is no longer normalized. 

These scaling results suggest a transition between a standard, Gaussian, random walk regime and a L\'evy regime as $\widetilde{U}_0$ goes below 5. For $\widetilde{U}_0> 5$, the variance of $\chi$ is finite [since $q(\chi)\sim |\chi|^{-3}$ when $\widetilde{U}_0=5$] and so, based on the central limit theorem argument presented in Sec.~\ref{Section:IntroLevy}, it is to be expected that a Gaussian position distribution of $P(x,t)$ will be obtained in the long time limit. For $\widetilde{U}_0<5$, on the other hand, the variance of $\chi$ diverges, suggesting a L\'evy flight scenario (at least as long as $\widetilde{U}_0>1$ and the mean flight-duration $\tau$ is finite). $P(x,t)$ attains the self-similar scaling form:
\begin{equation} 
  P(x,t) \sim \frac{1}{(K_{\nu} t)^{1/\nu}} L_{\nu,0} \left[ \frac{ x}{(K_{\nu} t)^{1/\nu} } \right]. 
  \label{eqLevy} 
\end{equation} 
Here $L_{\nu,0}(x)$ is the symmetric L\'evy stable PDF of Eq.~\ref{Eq:SymLevy}. The transport coefficient $K_{\nu}$ describes the width of the packet and is given in terms of the microscopic parameters of the model~\cite{Kessler2012}. The L\'evy exponent $\nu= ( \widetilde{U}_0 + 1) / 3$ is such that the solution approaches a Gaussian when $\widetilde{U}_0\to5$. $K_\nu$ vanishes as $\widetilde{U}_0\to 1$ due to the divergence of the average $\tau$. 

\begin{figure} 
  \centering
    \begin{overpic}
      [width=\linewidth]{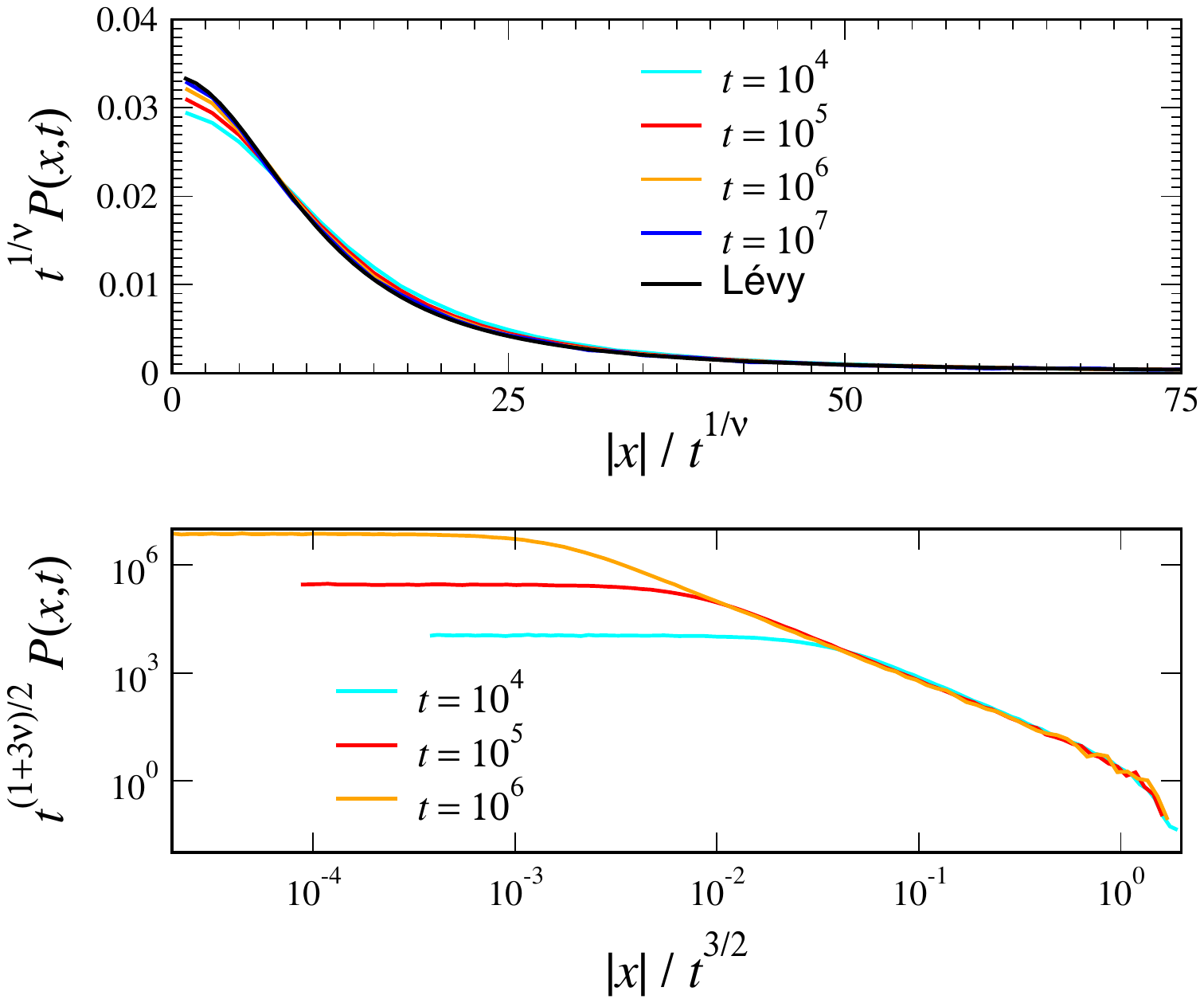}
    \end{overpic}
    \caption{Adapted from~\cite{Kessler2012}. Langevin simulations of the spatial dynamics. (Top) Results of numerical simulations of Eq.~\ref{eq:Langevin}. Packets of particles starting from a common origin converge, in the long time limit, to a self-similar L\'evy density presented as a solid line (Eq.~\ref{eqLevy}, $\nu=7/6$). (Bottom) The simulated distribution exhibits also a second scaling, \ie data collapse for several measurements times. The convergence is faster for large $x/t^{3/2}$ while for small values deviations from scaling are found. At large distances the power-law decay of the L\'evy density is cut off  due to finite time effects.}
	{\label{fig:fig6}}
\end{figure}

As discussed in Sec.~\ref{Section:IntroLevy}, power-law tails lead to infinite moments and hence the L\'evy law in Eq.~\ref{eqLevy} as a stand alone solution is not valid for very large $x$ and finite time. The resolution of this paradox lies in the fact that the random variables $\tau$ and $\chi$ are in fact correlated, since longer flight durations lead to larger displacements. In fact, the largest jump in the process $p(t)$ cannot be much larger than a length scale that increases as $t^{3/2}$, beyond which the tails of the L\'evy PDF are naturally cut off. This is made evident by completely neglecting the restoring friction force. Then the momentum undergoes pure diffusion, scaling like $t^{1/2}$ and the jump size scales accordingly as $t^{3/2}$. This is a kind of L\'evy walk, rather than a L\'evy flight. It is different from the original L\'evy walk, where the largest jump scales linearly with measurement time, since there the velocity is constant between turning points whereas here the motion is stochastic between any two zero crossings. This spatial regime, where the L\'evy density (Eq.~\ref{eqLevy}) holds, is therefore valid only up to a length scale that grows like $t^{3/2}$. This L\'evy regime also shrinks as $\widetilde{U}_0$ increases and vanishes as $\widetilde{U}_0\to 5$, beyond which only the Gaussian regime survives. To handle these correlations, a powerful tool called the Montroll-Weiss equation~\cite{MontrollWeiss} has to be employed~\cite{metzler2000random, Zaburdaev2015Review}, relating the joint PDF of jump distances and waiting times to the density of particles $P(x,t)$ using the convolution theorem of Laplace and Fourier transforms. Going through the analysis of the Montroll-Weiss equation, it is indeed found that Eq.~\ref{eqLevy} is valid in the central regime (Fig.~\ref{fig:fig6}).

The analysis of the far tail of $P(x,t)$ was carried out in~\cite{Aghion2017}, finding a spatial infinite (\ie non-normalized) density. It showed a relation between the laser cooling process and the problem of the distribution of random areas under Langevin excursions~\cite{Majumdar2005,Barkai2014,Agranov2020}. The latter is a constrained Langevin process starting and ending at $p=0$, never crossing the origin within a given time interval (Fig.~\ref{fig:fig5}). The derived cutoff of $P(x,t)$ due to the above-mentioned correlations, ending the power-law L\'evy regime, can be seen at the far right edge of the bottom panel of Fig.~\ref{fig:fig6}, and is in accord with direct simulations of the Langevin Eq.~\ref{eq:Langevin}. 

When $\widetilde{U}_0<1$, the correlations between the walk duration and distance can never be neglected and then $P(x,t)\sim (1/t^{3/2}) g(x/t^{3/2})$, where $g$ is some scaling function. In this limit of very shallow lattices, the momentum performs a random walk due to the random emission events, the friction is negligible and then the momentum scales like $\sqrt{t}$ as in Eq.~\ref{eqppp}. Hence, as shown analytically in~\cite{Barkai2014}, a cubic scaling of the MSD is obtained. This type of Richardson-like behavior, measured originally in~\cite{Richardson1926} following the distance between two weather balloons in a turbulent atmosphere, was found by~\cite{Wickenbrock2012} using Monte Carlo simulations and the phase $\widetilde{U}_0<1$ is called the Richardson phase. Richardson's law has thus far evaded measurement in atomic systems but has been experimentally observed in other contexts~\cite{Duplat2013}. 

The MSD therefore has three distinct phases. Normal diffusion, superdiffusion and Richardson diffusion:
\begin{equation} 
  \langle x^2 \rangle \sim \left\{ \begin{array}{c c} t \ \ \ & 5< \widetilde{U}_0 \\ t^{(7-\widetilde{U}_0)/2} \ \ & 1< \widetilde{U}_0 < 5\\ t^3 \ \ \ \ & \widetilde{U}_0 <1 . \end{array} \right. 
  \label{eq:xMSD} 
\end{equation} 
This behavior differs from the original L\'evy walk picture of Eq.~\ref{eq04} which exhibits at most ballistic spreading. An intuitive argument for the behavior of the MSD in the regime $1< \widetilde{U}_0 < 5$ can be obtained considering that the L\'evy distribution of Eq.~\ref{eqLevy} describes the central part of the packet and that a cutoff exists at distances of order $t^{3/2}$ (Fig.~\ref{fig:fig6}). From the power-law tail of the distribution, $P(x,t) \simeq t x^{- 1 - \nu}$ for $x<t^{3/2}$ can be obtained. Then $\langle x^2 \rangle \simeq \int^{t^{3/2}} dxx^2 P(x,t) $, and using $\nu= (\widetilde{U}_0 +1)/3$, the result of Eq.~\ref{eq:xMSD} is found. In short, the cutoff of the spatial L\'evy distribution gives the correct time dependence of the MSD, but to calculate the MSD precisely, including prefactors, one needs to resort to the scaling Green-Kubo theory investigated in Sec~\ref{Section:GreenKubo}.

\begin{figure} 
  \centering
    \begin{overpic}
      [width=\linewidth]{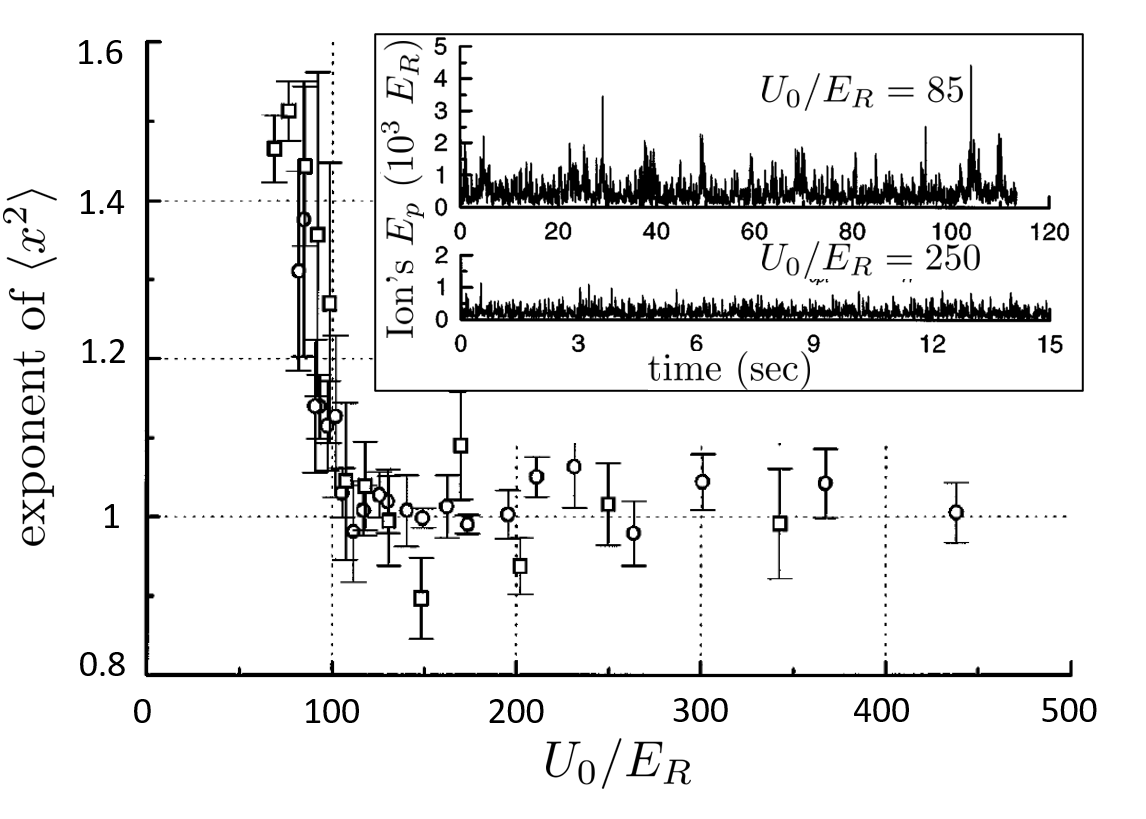}
    \end{overpic}
    \caption{Adapted from~\cite{Katori1997}. The exponent of the MSD $\langle x^2\rangle$ experimentally measured from the dynamics of a weakly harmonically trapped single $^{24}\mbox{Mg}^{+}$ ion in a Sisyphus lattice. For deep lattices, the dynamics are diffusive and $\langle x^2\rangle\sim t$. As the lattice becomes shallower a steep rise in the exponent is observed as the dynamics become superdiffusive (Eq.~\ref{eq:xMSD}). The inset shows the time traces of the potential energy of the ion as it moves in the trap. Large rare events are visible in the shallow lattice (inset, top) corresponding to the heavy tails of the distributions, compared to the more Gaussian behavior of the deep lattice data (inset, bottom).}
	{\label{fig:fig7}}
\end{figure}

The transition from normal to anomalous diffusion was observed experimentally in~\cite{Katori1997}, where the axial motion of a single $^{24}\mbox{Mg}^{+}$ ion trapped in a quadrupole ring trap undergoing one dimensional Sisyphus cooling by a pair of slightly red-detuned counter propagating laser beams was used to measure the MSD. The position of the ion was continuously measured by detecting its fluorescence photons through a high numerical aperture microscope objective to within a spatial resolution of 3 microns and temporal resolution of 10 microseconds. The scaling exponent of the MSD was observed to rise above unity below some threshold value of $\widetilde{U}_0$, continuing to rise with decreasing $\widetilde{U}_0$. The rise was roughly linear with $\widetilde{U}_0$, and the slope consistent with the theoretical prediction of unity. Fig.~\ref{fig:fig7} depicts the exponent of the MSD of the trapped ion, as well as its time-traces showing a clear transition from normal statistics for deep lattices to rare-event dominated statistics for shallow ones.

The spatial dynamics was further explored in~\cite{Sagi2012}, where the 1d diffusion of cold \Rb atoms undergoing Sisyphus cooling was studied. Starting with a narrow, thermally-equilibrated atomic cloud\footnote{This is achieved by allowing a long evaporation time where collisions thermalize the atomic cloud. The momentum distributions after the evaporation stage are verified to be Gaussian with variance corresponding to a temperature of 12~$\mu$K.}, the particles were released and their density profile absorption-imaged. In order to enable long-time measurements of the 1D dynamics and minimize the escape of atoms into other orthogonal dimensions, a far-detuned ``tube"-dipole trap was used with a geometry that generated strong confinement on the radial axes but a negligible effect on the experimental axis as defined by the Sisyphus lattice beams.

\begin{figure} 
  \centering
    \begin{overpic}
      [width=\linewidth]{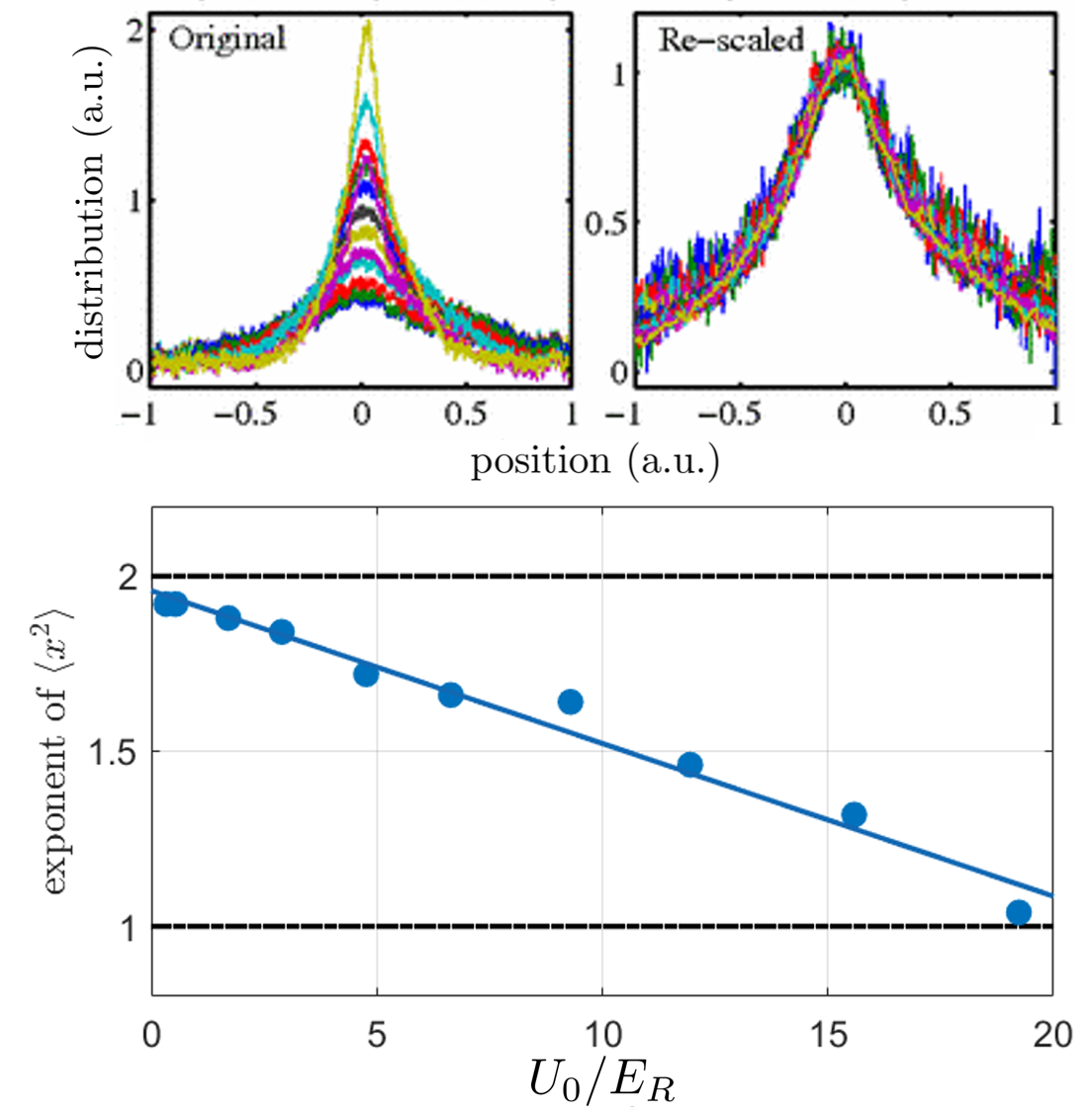}
    \end{overpic}
    \caption{Adapted from~\cite{Sagi2012}. Self-similarity, L\'evy distributions and superdiffusion. The temporal scaling of the spatial distribution is experimentally measured in 1d. (top) Data collapse of the spatial distributions as a function of time for the rescaling transformation $x\to x/t^{1/\nu}$ (Eq.~\ref{eqLevy} and Fig.~\ref{fig:fig6}), with $\nu = 1.25$ in a $U_0/E_R = 4.8$ lattice. (bottom) The extracted exponent of the MSD $\langle x^2\rangle$. $C$ of Eq.~\ref{eq:defU0} is used as a fitting parameter for Eq.~\ref{eq:xMSD}, giving a value of $11\pm1$ for this measurement. The experiment also observed the power-law tails of the spatial distribution predicted by Eq.~\ref{eqLevy}.}
	{\label{fig:fig8}}
\end{figure}

In qualitative agreement with theory, a transition from normal to anomalous dynamics was observed and L\'evy distributions were found to fit well to the experimental data, confirming the transition between the normal Gaussian diffusion regime for deep enough lattices and the L\'evy regime below some critical $\widetilde{U}_0$. Both the power-law scaling of the MSD and the L\'evy distribution of the displacement, with the L\'evy index changing with lattice depth were observed, in agreement with theory. The scaling collapse and the resultant L\'evy distributions are shown in the top panel of Fig.~\ref{fig:fig8}. The bottom panel shows the measured exponent of the time-dependent MSD, as a function of $U_0/E_R$. The superdiffusive nature of the spatial spreading is also in agreement with the theoretical expectations and, as well, the linear dependence of the exponent with $U_0/E_R$ is in agreement with Eq.~\ref{eq:xMSD}. 

This agreement between theory and experiment, while encouraging, still remains incomplete. In particular, exponents above 2 and the saturation at a value of 3 associated with the Richardson phase have yet to be observed. This is not totally unexpected, since if the particles are moving super-ballistically they will quickly leave the trap and may go undetected~\cite{Sagi2012}. A possible way to alleviate this may arise in the form of optically-engineered potentials which create strong confinement as well as reflecting boundary-conditions for the atomic packet~\cite{Gaunt2013,Livneh2018}.

\subsection{Position-momentum correlations}
\label{Section:Correlations}
As both the momentum (Eq.~\ref{eqppp}) and position (Eq.~\ref{eq:xMSD}) dynamics are governed by power-laws, the cross-correlation $C_{xp}$ between position and momentum may also be expected to behave in a similar way. Defined according to
\begin{equation}
  C_{xp}(t)=\frac{\braket{x(t)p(t)}}{\sqrt{\braket{x^2(t)}\braket{p^2(t)}}},
  \label{eq:correlation_definition}  
\end{equation}
this function asymptotically decays as $1/\sqrt{t}$ for normal diffusion~\cite{Gillespie2012} and approaches unity for ballistic motion. This observable is in general challenging to access experimentally as it requires knowledge of the position of a group of atoms contained in a certain narrow momentum bin. 

\begin{figure}
	\centering
	\begin{overpic}
		[width=\linewidth]{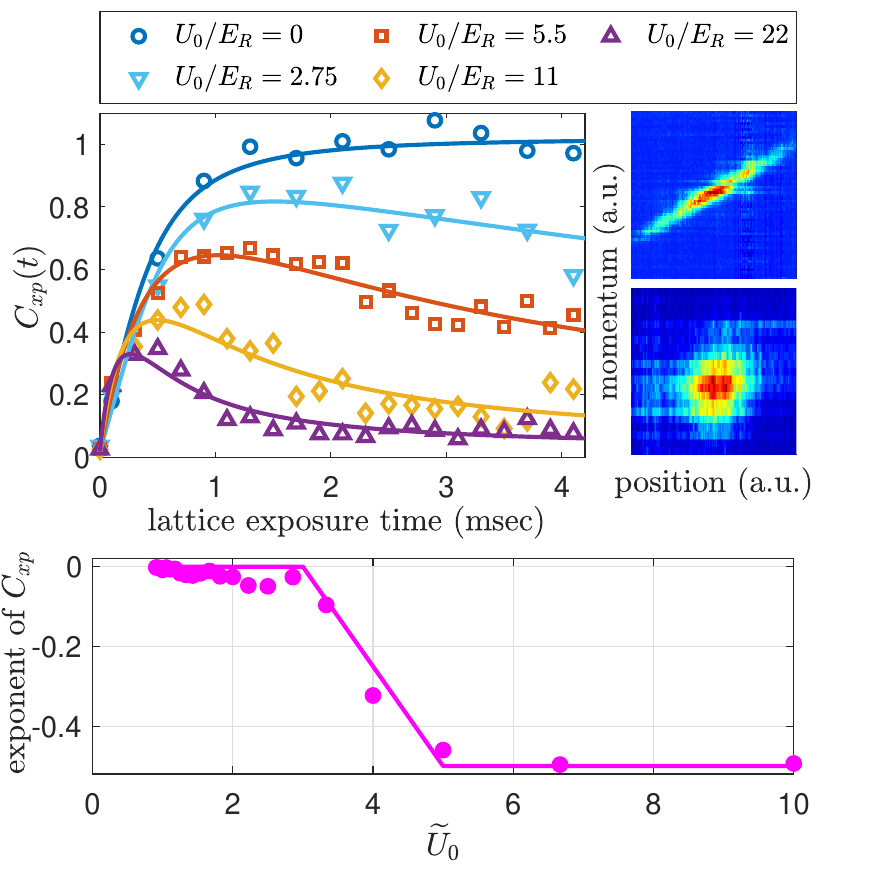}
	\end{overpic}
	\caption{Adapted from~\cite{Afek2017Correlations}. Anomalous dynamics in momentum and position lead to intriguing position-momentum correlations. (Top) Experimental results for the position-momentum correlations as a function of lattice exposure time and lattice depth $U_0/E_R$. At short times the correlations build up and are later quenched at varying rates, depending on the anomalous dynamics. The images on the right are the tomographically-measured phase space probability densities for ballistic expansion (upper) and $U_0/E_R = 5.5$ (lower), at a lattice exposure time of 4.1~msec, showing the effect of the lattice on the development of the correlations. (Bottom) Langevin simulation (circles) validating the prediction of Eq.~\ref{eq:scaling} depicted by the solid line.}
	\label{fig:fig9}
\end{figure}

In~\cite{Afek2017Correlations}, a scaling relation was derived describing the asymptotic dynamics of $C_{xp}(t\to\infty)$ for the general case where both position and momentum have power-law long-time behavior: $\braket{x^2(t)}\sim t^{2\alpha_x}$ and $\braket{p^2(t)}\sim t^{2\alpha_p}$. Here the exponents $\alpha_i$ describe anomalous processes in general. Substituting those into Eq.~\ref{eq:correlation_definition}, along with the fact that $\langle xp\rangle \sim d\langle x^2\rangle/dt\sim t^{(2 \alpha_x-1)}$, gives $C_{xp}\sim t^{\alpha_x-\alpha_p-1}$. In terms of $\widetilde{U}_0$, it should then scale like
\begin{equation}
  C_{xp}(t) \sim\left\{ \begin{array}{l l} \mbox{const} & \widetilde{U}_0<3 \\ t^{(3-\widetilde{U}_0)/4} & 3<\widetilde{U}_0<5 \\ t^{-1/2} & \widetilde{U}_0>5.\end{array} \right. \label{eq:scaling}
\end{equation}
As $\widetilde{U}_0$ is varied, the behavior ranges from ``normal", a $t^{-1/2}$ decay, to a constant. This has now been numerically verified in a new analysis of the simulations performed in~\cite{Afek2017Correlations} and is shown in the bottom panel of Fig.~\ref{fig:fig9}.

The correlation function is however not merely a tool to explore the long-time dynamics. Rather, it yields information about the short-time behavior as well. An additional feature discussed in~\cite{Afek2017Correlations} relates to the short-time dynamics of a system of particles released from a harmonic trap and allowed to propagate in the Sisyphus lattice. At times that are short compared to the oscillation period in the trap before release, ($t<1/\omega$), correlations build up linearly regardless of the depth of the lattice as the diffusive dynamics do not yet affect the atoms ($\omega$ sets, in accordance with the equipartition theorem, the ratio between the initial standard-deviation of the momentum distribution and that of the position distribution). At longer times, diffusion kicks in and the behavior of Eq.~\ref{eq:scaling} is expected to take over.

A setup similar to the one described in~\cite{Sagi2012} was used to generate the anomalous dynamics, and a method based on velocity-selective two-photon Raman transitions~\cite{Moler1992} was developed to tomographically image the phase-space density distribution function. In this method, atoms contained within a narrow velocity class are selectively transferred from the $\ket{F=1}$ lower hyperfine state to the upper $\ket{F=2}$ state using a Raman $\pi$-pulse of a given detuning. The center of the selected velocity class is scanned by varying the the two-photon detuning of the pulse, and the Rabi frequency sets its width~\cite{Kasevich1991}. The position of the selected atoms is then directly imaged using state-selective absorption imaging. This way, a direct measurement of the position-momentum correlation function is enabled for different initial conditions given by the Sisyphus lattice exposure. The two right panels of Fig.~\ref{fig:fig9} are experimental reconstructions of the phase space density distributions. The upper panel reveals a high correlation corresponding to ballistic expansion while the lower one shows the destruction of the correlation by the diffusive dynamics.

Fig.~\ref{fig:fig9} (top panel) shows the dynamics of these correlations as an atomic cloud evolves in lattices of different $\widetilde{U}_0$. The initial buildup of the correlation is evident for all lattice depths considered, as well as its following decay. The $\widetilde{U}_0=0$ data set is expected (and shown) to be ballistic.   

\begin{figure} 
  \centering
    \begin{overpic}
      [width=0.9\linewidth]{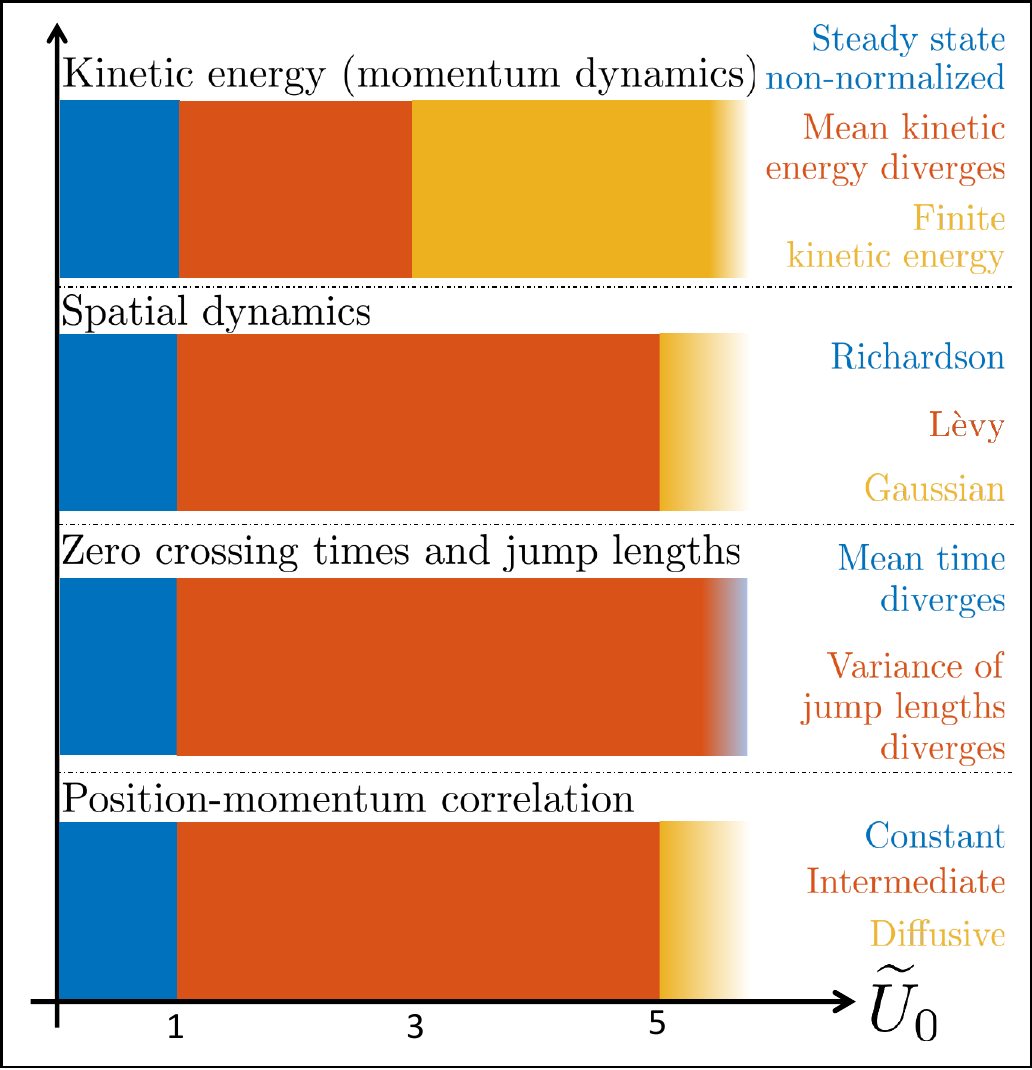}
    \end{overpic}
    \caption{A recap of the dynamical phases for various observables. Varying the experimentally-accessible control parameter $\widetilde{U}_0$, defined in Eq.~\ref{eq:defU0} and Fig.~\ref{fig:fig2} reveals transitions in the statistical properties of the system discussed in Section~\ref{Section:MomentumRealCorr}. Momentum, position and position-momentum correlations all display sharp transitions between different regimes. Nonphysical divergences are resolved when moments are calculated using the formalism of the infinite covariant density.}
	{\label{fig:fig10}}
\end{figure}

{\bf In summary,} The various aspects discussed in this section can be summarized in the form of a ``phase diagram". Fig.~\ref{fig:fig10} presents the dependence of the momentum (Eq.~\ref{eqppp}) and spatial dynamics (Eq.~\ref{eq:xMSD}) on the lattice depth parameter $\widetilde{U}_0$, defined in Eq.~\ref{eq:defU0}, as well as changes in the statistical properties of momentum zero-crossing times and jump distances (Eq.~\ref{eqZoller}) and even position-momentum correlation (Eq.~\ref{eq:scaling}). All of these display sharp transitions between various regimes as $\widetilde{U}_0$ is scanned.

\section{Implications for fundamental concepts in Statistical Physics}
\label{Section:FundamentalConcepts}

\subsection{Scaling Green-Kubo relation}
\label{Section:GreenKubo}

As mentioned in Sec.~\ref{Sec:Introduction}, the Green-Kubo relation which first appeared in~\cite{Taylor1922} in the context of diffusion in a turbulent medium, relates the diffusion constant of a particle, $D = \langle x^2(t)\rangle/2t$, to an integral over the stationary time correlation function of the momentum $C_p(t,t+\tau) = \langle p(t)p(t+\tau)\rangle$~\cite{Green1954,Kubo1957}. For a particle with mass $M$,

\begin{equation}
  D = \frac{1}{M^2}\int\limits_0^\infty d\tau C_p(t,t+\tau).
  \label{eq:GreenKubo}
\end{equation}

In the case of a Brownian particle with Stokes’ friction coefficient $\gamma_s$, the momentum correlation function $C_p(t,t+\tau) = k_B T M e^{-\gamma_s \tau}$ is exponential. The Green-Kubo formula gives the Einstein relation $D= k_B T/(M\gamma_s)$ linking the diffusivity and the friction coefficient via the temperature $T$. In the Sisyphus-cooled atomic system at hand, for $\widetilde{U}_0>3$, this standard Green-Kubo relation applies. For $\widetilde{U}_0<3$, however, it breaks down owing to one of two reasons: either because the correlation function is not stationary, or alternatively, while the correlation function is stationary, its time-integral diverges. Depending on the value of $\widetilde{U}_0$, either scenario can occur and in both cases Eq.~\ref{eq:GreenKubo} needs to be generalized. This is done following~\cite{Dechant2014}.

Consider the general momentum correlation function
\begin{equation}
  C_p(t_2,t_1)=\int\limits_{-\infty}^{\infty} dp_2\int\limits_{-\infty}^{\infty}dp_1 p_2 p_1 P(p_2,t_2| p_1,t_1) P(p_1,t_1|0,0)
  \label{eq:Corr}
\end{equation}
where $P(p_2,t_2|p_1,t_1)$ is the probability of the particle having momentum $p_2$ at time $t_2$, given that it had momentum $p_1$ at time $t_1$. The notation $P(p_2,t_2|p_1,t_1)$ is used here, as opposed to the $W(p,t)$ used earlier, to explicitly indicate the dependence on the initial momentum and time. \braket{ x^2(t)} can then be calculated from $C_p(t_2,t_1)$. In the case where the process is stationary at long times, so that the stationary momentum distribution (denoted by the subscript $\textrm{s}$) exists and is given by $\lim_{t_1\to \infty} P(p_1,t_1|0,0) = W_{\rm{s}}(p_1)$, then $C_p(t_2,t_1)=C_{p,\textrm{s}}(|t_2-t_1|)$ only depends on the time lag $\tau=t_2-t_1$
\begin{equation}
C_{p,\textrm{s}}(\tau) = \int\limits_{-\infty}^{\infty} dp_2\int\limits_{-\infty}^{\infty} dp_1 p_2 p_1 P(p_2,\tau| p_1,0) W_\textrm{s}(p_1)
\end{equation}
Since $x(t)$ is the time-integral of $p(t)/M$, 
\begin{align}
  \langle x^2(t)\rangle &= \frac{1}{M^2} \int\limits_0^t dt_2 \int\limits_0^t dt_1 C_{p,\textrm{s}}(|t_2-t_1|)\nonumber\\
  &=\frac{2t}{M^2}\int\limits_0^\infty d\tau C_{p,\textrm{s}}(\tau)
\end{align}
from which follows the standard Green-Kubo formula, Eq.~\ref{eq:GreenKubo}.

What then is the form of the momentum correlator $C_p(t_2,t_1)$ in the Sisyphus system? Turning to the definition of Eq.~\ref{eq:Corr}, we have already seen how to calculate from the Fokker-Planck Eq.~\ref{eqLutz} the factor $W(p_1,t_1)$. For large $p_1$, this decays as a power-law with a Gaussian fall-off at $p_1 \sim O(\sqrt{t_1})$ (Eqs.~\ref{eq:Wscaling},~\ref{eq:Wsolf} and \ref{eq:Wsolfasym}). The calculation of the other factor $P(p_2,p_1;t_2-t_1)$ is similar and it behaves similarly for large $p_2$ as long as $p_1$ is not too large which is the relevant regime since the $P(p_1,t_1|0,0)$ factor in Eq.~\ref{eq:Corr} cuts it off. The detailed calculation~\cite{Dechant2011,Dechant2012a} reveals that the correlator has the following form:
\begin{equation}
  C_p(t_2,t_1) \approx \left\{\begin{array}{ll}
  {\cal C}_{\alpha>1} t_1^{2-\alpha}g_{\alpha>1}\left(\frac{t_2-t_1}{t_1}\right), 
  &\alpha>1\\
  {\cal C}_{\alpha<1}t_1 g_{\alpha<1}\,\left(\frac{t_2-t_1}{t_1}\right),
  &\alpha<1\end{array}\right.
\end{equation}
where $\alpha \equiv \frac{\widetilde{U}_0+1}{2}$, and
\begin{align}
  {\cal C}_{\alpha>1}&=\frac{{\cal N}(4D_0/p_c)^{2-\alpha}\sqrt{\pi}}{\Gamma(\alpha+1)\Gamma(\alpha)/p_c}\nonumber\\
  {\cal C}_{\alpha<1}&=\frac{4D_0\sqrt{\pi}}{\Gamma(\alpha+1)\Gamma(1-\alpha)} \nonumber\\
  g_{\alpha>1}(s)&=s^{2-\alpha}\int\limits_0^\infty dy\,y^2 e^{-y^2}{}_1F_1\left(\frac{3}{2};\alpha+1;y^2\right)\Gamma\left(\alpha,y^2s\right)\nonumber\\
  g_{\alpha<1}(s)&=s\int\limits_0^\infty dy\, y^2 e^{-y^2}{}_1F_1\left(\frac{3}{2};\alpha+1;y^2\right)e^{-y^2s}.
\end{align}
${}_1F_1$ is the confluent hypergeometric function, and ${\cal N}$ is the normalization factor given below Eq.~\ref{eq07}. The correlation function is, in general, non-stationary. The dependence of the correlator on the relative time $s=(t_2-t_1)/t_1$ is called ``aging". Normally the aging correlator has the form $\langle C_p(t_2,t_1)\rangle \sim \langle p^2\rangle_\textit{eq}g(s)$~\cite{Bouchaud1992,Margolin2004,Burov2010}, however here it takes the form
\begin{equation}
 C_p(t_2,t_1)
\sim t_1^\phi g(s)\sim\langle p^2(t_1)\rangle g(s),
\label{eq:superage}
\end{equation}
with $\phi=\min(2-\alpha,1)$, and $g(s)$ is either $g_{\alpha>1}$ or $g_{\alpha<1}$ depending on whether $\alpha$ is greater or smaller than unity. To mark this added dependence on $t_1$ due to the growth of $\langle p^2(t_1) \rangle$ with time, this phenomenon is termed ``super-aging". 

Let us consider the limit $t_2-t_1\ll t_1$,
\begin{equation}
C_p(t_2,t_1) \approx \left\{ \begin{array}{ll}\frac{\pi\Gamma(\alpha-2)}{4{\cal N}\Gamma^2\left(\alpha-\frac{1}{2}\right)}\left[4D_0(t_2-t_1)\right]^{2-\alpha}&\alpha>2\\
\frac{1}{{\cal N}\Gamma(\alpha)\Gamma(2-\alpha)}\left(4D_0t_1\right)^{2-\alpha} & 1<\alpha<2 \\
(1-\alpha)(4D_0t_1)&\alpha<1\end{array}\right.
\end{equation}
For $\alpha>2$, $\phi=2-
\alpha$ and the $t_1^\phi$ factor cancels against the $t_1$ factor in $g(s)$, leaving the correlation function stationary in this limit, a fact that will be important for the discussion of ergodicity breaking in Sec.~\ref{Section:Ergodicity}. For $\alpha<2$ on the other hand, even the limiting correlation function is nonstationary and dominated by the growth in time of $\langle p^2(t_1)\rangle$, leading to super-aging.  

The super-aging of the correlation function (Eq.~\ref{eq:superage}) has some interesting consequences on the MSD~\cite{Dechant2014}. For $t\gg 1$,
\begin{align}
  \langle x^2(t)\rangle &\simeq \frac{2{\cal C}_{f}}{M^2}\int\limits_0^t dt_2\int\limits_0^{t_2} dt_1\, t_1^\phi g\left(\frac{t_2 - t_1}{t_1}\right)\nonumber\\
  &\simeq \frac{2{\cal C}_{f}}{M^2}\int\limits_0^t dt_2\, t_2^{\phi+1} \int\limits_0^\infty ds (s+1)^{-\phi-2}g(s)\nonumber\\
  &\simeq 2D_\phi t^{\phi+2}
  \label{eq:GenGK}
\end{align}
with
\begin{equation}
  D_\phi\equiv \frac{{\cal C}_{f}}{M^2(\phi+2)}\int\limits_0^\infty ds\, (s+1)^{-\phi-2}g(s).
\end{equation}
Here, ${\cal C}_f$ is either ${\cal C}_{\alpha>1}$ or ${\cal C}_{\alpha<1}$, as appropriate. This reproduces the scaling behavior of Eq.~\ref{eq:xMSD} on general grounds from the L\'evy scaling and the cutoff, adding to it the calculation of $D_\phi$. 

Eq.~\ref{eq:GenGK} is the scaling form of the Green-Kubo relation. It is applicable for $\phi>-1$ which corresponds to superdiffusion. The usual diffusion coefficient in Eq.~\ref{eq:GreenKubo} is then ill-defined, and is replaced by the anomalous diffusion coefficient $D_\phi$. As with the original Green-Kubo formula, $D_\phi$ is given in terms of an integral over a function of a single variable. Determining the diffusive behavior of a system from its correlation function thus amounts to determining the exponent $\phi$ and the scaling function $g(s)$. While Eq.~\ref{eq:GenGK} was derived in terms of momentum and position, it holds for any two quantities where one is up to a constant factor the time integral of the other. An example of such analogy, between frequency and phase, will be given in Sec.~\ref{Section:MotionalBroadening}. 

The different scaling regimes for the MSD for the Sisyphus problem are thus seen to be related to the properties of the correlator. For $\alpha>3$ ($\widetilde{U}_0>5$), the correlator is stationary, the integral in the standard Green-Kubo formula converges and the diffusion is normal. The spatial diffusion constant diverges as $\alpha\to 3$ from above, due to the factor $s^{2-\alpha}$ in $g$, signalling the breakdown of normal diffusion. For $1<\alpha<3$ ($1<\widetilde{U}_0<5$), the standard Green-Kubo formula breaks down and the exponent $\phi=2-\alpha$. The MSD then scales as $t^{4-\alpha}$ and the dynamics are superdiffusive, with an anomalous diffusion constant given by (plugging in the value of ${\cal N}$) 
\begin{equation}
  D_\phi=\frac{(4D_0)^{2-\alpha}p_c^{2\alpha-2}\Gamma(\alpha-1/2)(\alpha-1)}{\Gamma^3(\alpha)\alpha(4-\alpha)M^2}\int\limits_0^\infty ds\, (s+1)^{\alpha-4}g_{\alpha>1}(s).
\end{equation}
This vanishes at $\alpha=1$ ($\widetilde{U}_0=1$) and, due to the factor $s^{2-\alpha}$ in $g_{\alpha>1}(s)$, diverges as $\alpha\to 3$. 

For $\alpha<1$ ($\widetilde{U}_0<1$) the anomalous diffusion exponent saturates at a value of 3, corresponding to Richardson diffusion. The anomalous diffusion constant in this regime is
\begin{equation}
D_\phi=\frac{4D_0\sqrt{\pi}}{3\Gamma(\alpha)\Gamma(1-\alpha)M^2}\int\limits_0^\infty ds\, (s+1)^{-3} g_{\alpha<1}\left(s\right).
\end{equation}

The momentum correlation function can be directly measured using existing experimental techniques similar to those described in Sec.~\ref{Section:Correlations}. For example, an extremely narrow atomic momentum distribution centered around a specifically targeted momentum $p_0$ can be prepared using two-photon momentum selective Raman transition with two counter-propagating, far-detuned laser beams~\cite{Kasevich1991} and then exposed to Sisyphus cooling for a variable amount of time $t_1$. The resulting momentum distribution $P(p)$ can then be measured again after an additional time $t_2$ with high resolution using the same Raman momentum selection method, directly yielding the correlation function $C_p(t_2,t_1)$ and through it the Green-Kubo relation. 

\subsection{Breakdown of ergodicity}
\label{Section:Ergodicity}

Systems in equilibrium visit all of phase space, with the average relative frequency of visiting any certain point given by the Boltzmann-Gibbs distribution. Thus, over long enough observation times, the time-averages of observables correspond to the equilibrium ensemble averages. In systems with anomalous diffusion, however, this is no longer necessarily the case~\cite{metzler2014anomalous}. The time-average, even in the infinite time limit, varies from realization to realization. This was studied in the context of sub-recoil laser cooling~\cite{Saubamea1999}, fluorescence intermittency in quantum dots~\cite{Brokmann2003}, and single-atom motion in non-dissipative optical lattices~\cite{Kindermann2016Nature}. Ergodicity breaking in Sisyphus cooling was first theoretically investigated in~\cite{Lutz2004}. An alternative analysis taking into account the essential time-dependence of the momentum distribution at large momenta is given in~\cite{Dechant2011,Dechant2012a}.

To probe the possibility of ergodicity breaking for the momentum, consider the ensemble variance of the difference between the time average $\bar{p}$ and the ensemble average $\langle p \rangle$ for a particle that starts with momentum $p=0$ at $t=0$. Ergodicity is broken when $\bar{p}-\langle p \rangle$, which reduces to  
\begin{equation}
 \langle {\bar{p}}^2 \rangle = \lim_{t\to\infty} \frac{1}{t^2}\langle{} x^2\rangle
\end{equation}
for symmetric distributions, deviates from zero. Given Eq.~\ref{eq:xMSD} describing the spatial MSD,
we find~\cite{Dechant2012a}
\begin{equation}
 \langle \bar{p}^2 \rangle \sim \left\{ \begin{array}{ll} t^{-1} & \alpha>3\\
  t^{2-\alpha} & 1<\alpha<3 \\
  t & \alpha<1. \end{array}\right.
\end{equation}
The $1/t$ behavior seen for $\alpha>3$ ($\widetilde{U}_0>5$) is the ``normal" behavior. For $2<\alpha<3$ ergodicity is achieved, albeit anomalously slower. For $\alpha<2$, $\langle \bar{p}^2\rangle$ does not vanish as $t\to\infty$ and ergodicity is broken. 

\subsection{Violation of the equipartition theorem}
\label{Section:Equipartition}

The previous sections focused on free, untrapped systems [with the exception of the trapped ion system of Fig.~\ref{fig:fig7}~\cite{Katori1997}]. It is natural to now address the effect of an external binding potential, whose scale is much larger than the wavelength of the Sisyphus lattice and is hence not averaged out in the semiclassical treatment. Generally, a violation of equipartition in this inherently non-equilibrium system might be expected. The question that then arises is ``can the effect be quantified?" In this case, instead of simply searching for an expected violation of equipartition, the issue becomes one of determining the relation between moments of position and momentum for the non-linear friction under study. These have been studied theoretically, both in the simple context of a harmonic well~\cite{Dechant2015PRL,Dechant2016} and recently in a more general context~\cite{Falasco2022}. These works focus on the steady-state phase-space distribution of the particles and in particular the breakdown of equipartition and the virial theorem. The dynamics and violation of the equipartition of the energy in the system have also been probed experimentally in~\cite{Afek2020}. 

Deviations from equipartition in a harmonic potential can be parametrized using the \emph{equipartition parameter}- the square-root of the ratio of the expectation value of the kinetic energy $E_k$ to that of the potential energy $E_p$,
\begin{equation}
  \chi_H = \sqrt{\frac{E_k}{E_p}} =\sqrt{\frac{\langle p^2 \rangle}{\omega^2\langle x^2 \rangle}}. \label{eq:chi}
\end{equation}
The relevant parameters for the description of the system are $\mathcal{A}$, the strength of the linear friction, $p_c$, the momentum scale for the onset of the nonlinear friction (Eq.~\ref{eq06}), the diffusion constant $D_0$ and $\omega$, the harmonic oscillation frequency of the atoms in the trap. All but the latter depend experimentally on the detuning and intensity of the Sisyphus lasers. Consider the scaled coordinates $x \to M\omega x/p_c$ and $p\to p/p_c$. There are now two dimensionless parameters which characterize the dynamics, the by now familiar $\widetilde{U}_0$ relevant to the potential-free problem and a new parameter controlling the strength of the harmonic trap, $\Omega=\omega/\mathcal{A}$. The Fokker-Planck equation is modified to a Kramers-Fokker-Planck (KFP) equation for the phase-space density, which in these scaled coordinates reads:
\begin{align}
\begin{split}
  &\pd{}{t} P(x,p,t) =\\
  &\left[\Omega\left(-p\pd{}{x} + x\pd{}{p}\right) + \pd{}{p}\left(\frac{p}{1+p^2} + \frac{1}{\widetilde{U}_0}\pd{}{p}\right)\right]P(x,p,t).
\end{split}
\end{align}

The theoretical analysis focuses on the properties of the steady-state solution, which breaks down for shallow lattices so that a steady-state momentum distribution does not exist. An analysis of the time-dependent problem which is required to treat $\widetilde{U}_0<1$, has not yet been performed. 

The steady-state of the KFP equation is not exactly solvable, and hence the analysis is restricted to various limits and numerics~\cite{Dechant2015PRL,Dechant2016}. In short, the numerics predict that $\chi_H$ decreases for increasing lattice depth $\widetilde{U}_0$, dropping to an minimum value that depends on $\Omega$. This is followed by an asymptotic ascent back to unity as the lattice depth increases further. 

Experimentally observing such a violation of equipartition requires the ability to measure $\braket{x^2}$, $\braket{p^2}$ and $\omega$ according to Eq.~\ref{eq:chi}. In~\cite{Afek2020}, a setup similar to that described in~\cite{Afek2017Correlations} was used in conjunction with a superimposed crossed-optical dipole trap providing the confining potential. The atoms were coupled to a 1d dissipative Sisyphus lattice for a given amount of time, after which their spatial and momentum distributions were imaged using direct absorption imaging and time-of-flight respectively. Imaging the atomic density profile in-situ is susceptible to imaging errors arising from the high density of the trapped atoms, and avoiding this required the authors to homogeneously excite a controlled fraction of the atoms to the upper hyperfine state using a variable-length microwave pulse, and selectively image this transferred population. This provided a knob for scanning the density of the atoms and a way of extrapolating it down to zero giving the true size of the atomic cloud in an aberration-free way. The oscillation frequency was measured independently by applying a perturbation to the trapped cloud and observing its sloshing oscillations in the trap. 

\begin{figure} 
  \centering
    \begin{overpic}
      [width=\linewidth]{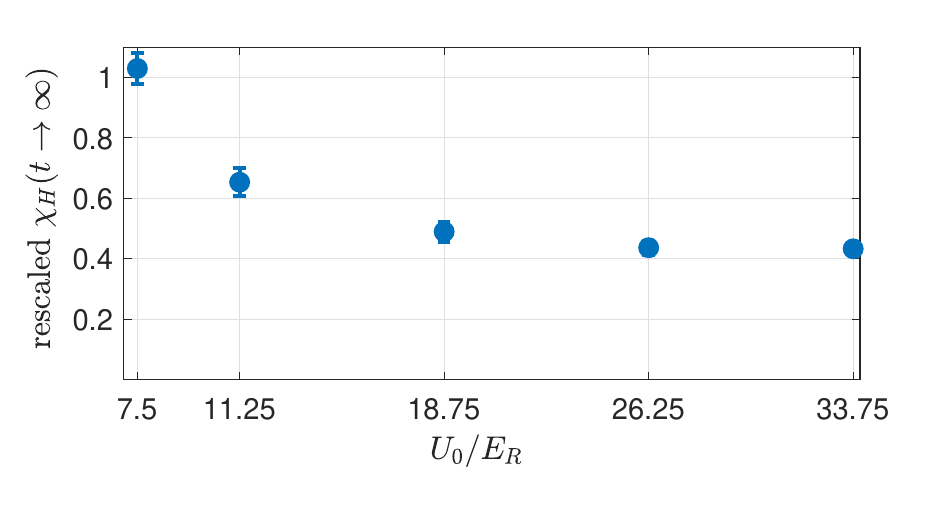}
    \end{overpic}
    \caption{adapted from~\cite{Afek2020}. Fundamental concepts such as the equipartition theorem are shown to be violated. Steady-state values of the equipartition parameter $\chi_H$ (Eq.~\ref{eq:chi}) experimentally measured for a trapped cloud of \Rb atoms (rescaled by its thermal value which is $< 1$ due to trap anharmonicity), quantifying a non-monotonic deviation from equipartition.} 
  \label{fig:fig11}
\end{figure} 

The measured steady-state values of $\chi_H$ are shown in Fig.~\ref{fig:fig11}. They display a monotonic decrease in $\chi_H$ as the lattice depth increases, perhaps hinting to the shallow-lattice behavior numerically observed in~\cite{Dechant2016}. The work carried out in~\cite{Afek2020} suffered greatly from the anharmonicity of the optical trapping potential. This necessitated the use of several non-trivial experimental methods to extract the harmonic $\chi_H$. 

To summarize, both theory and experiment exhibit violation of the equipartition theorem.
More refined experiments and possibly extension of the semi-classical approach are needed to quantify the effects more precisely.


\section{L\'evy statistics and power-laws in other atomic systems}
\label{Section:OtherSystems}

The generality of the framework presented in this Colloquium can be appreciated by observing other atomic systems, even when instead of momentum and position one considers a different pair of variables, \eg phase and frequency. This has profound implications for coherence times of quantum memories. Furthermore, the question of the relation of infinite ergodic theory with other laser cooling mechanisms is also addressed.

\subsection{Motional broadening in two-level system ensembles with a heavy-tailed frequency distribution}
\label{Section:MotionalBroadening}

Consider an oscillator whose frequency has some anomalous stochastic dynamics. This fluctuating frequency, together with its time integral, namely the phase $\phi$ of the oscillator, are analogous to the momentum and position whose dynamics were discussed above. The distribution of phases of an ensemble of oscillators spreads, similarly to the spatial spreading of the particle packet, leading to decoherence, which is a limiting factor in many applications such as atomic clocks and quantum memories based on two-level systems~\cite{Ludlow2015,Heshami2016}. In the typical case, when the instantaneous fluctuating frequency distribution of the ensemble has finite moments, stochastic fluctuations cause the phase to spread diffusively, $\Delta\phi\sim t^{1/2}$, as compared to a ballistic spread $\Delta\phi\sim t$ for a static frequency inhomogeneity~\cite{Sagi2010}. This slower diffusive spread induces the well-known effect of \emph{motional narrowing} of the power spectrum, which also lengthens ensemble coherence times~\cite{Bloembergen1948,Dicke1955}. When the instantaneous  frequency distribution of the ensemble is heavy-tailed, however, the picture is different. The stochastic phase dynamics becomes anomalous, the phase spread grows super-diffusively, and motional narrowing is hindered~\cite{Sagi2011}. In particular, when the first moment of the frequency distribution diverges, the stochastic frequency fluctuations can lead to \emph{broadening} of the spectrum (motional broadening), surprisingly shortening the coherence time as the rate of fluctuations increases~\cite{Sagi2011,BURNSTEIN1981335}. For a frequency distribution following L\'evy statistics $\sim\exp\left(-A|\kappa|^\nu\right)$ as in Eq.~\ref{eq03}, the transition between motional narrowing and motional broadening occurs at $\nu=1$, corresponding to the Lorenzian spectrum. The coherence time of the ensemble decays as $\tau_c^{\nu-1}$~\cite{Sagi2011} with $\tau_c$ being the correlation time of the fluctuating frequency. This slowing down or acceleration of the ensemble coherence decay, depending on the value of $\nu$, is a particularly striking feature. In this respect, the fluctuations act as resetting events making motional narrowing analogous to the Zeno effect~\cite{Milburn1988} in which certain events, such as measurements, delay the decay of a system. By the same token, motional broadening is analogous to the anti-Zeno effect~\cite{Sagi2011}, where the opposite occurs. 

It has been theorized~\cite{Poletti2012,Poletti2013Strong}, and recently verified experimentally~\cite{Bouganne2020} with ultracold atoms in optical lattices, that strong interactions in a many-body system can also generate an anomalous decay of the coherence of the ensemble. Long-range interactions in ion chains have recently been used to probe the assumption that classical hydrodynamics can emerge universally for any complex quantum system, due to mixing of local degrees of freedom through Ising interactions with a cleverly-engineered power-law decay~\cite{joshi2021observing}. Power-law spectral line shapes can naturally emerge in NMR due to dipolar interactions~\cite{klauder1962spectral}, long-range interactions~\cite{holtsmark1919verbreiterung}, spread in activation energies~\cite{Yue2016} or hopping distances~\cite{Paladino2014}, and various other homogeneous imperfections~\cite{stoneham1969shapes}. They are also related to ergodicity breaking in blinking quantum dots~\cite{MargolinPRL2005} and to the dynamics of photons in warm atomic vapor~\cite{Mercadier2009,Baudouin2014}.

\subsection{L\'evy dynamics in sub-recoil laser cooling}
\label{Section:SubRecoil}

Sub-recoil laser cooling~\cite{Aspect1988,Kasevich1992} relies on a totally different mechanism than that of the sub-Doppler cooling presented in Sec.~\ref{Section:IntroMechanism}, and still it is rewarding that some of the general insights gained by analyzing one can be applied to the other, showing the generality of the toolbox presented in this Colloquium. In particular, L\'evy laws are known to govern the statistical aspects of the problem, an issue previously analyzed extensively by Cohen-Tannoudji and collaborators~\cite{Bardou2002}. However, the role of infinite ergodic theory and its associated non-normalized state was only very recently connected to this system~\cite{barkai2021transitions}. The new analysis shows that in sub-recoil laser cooling there exists a non-normalised state describing some of the coldest atoms, that is complementary to the standard description~\cite{Bardou1994}. 

\begin{figure} 
    \centering
        \begin{overpic}
            [width=\linewidth]{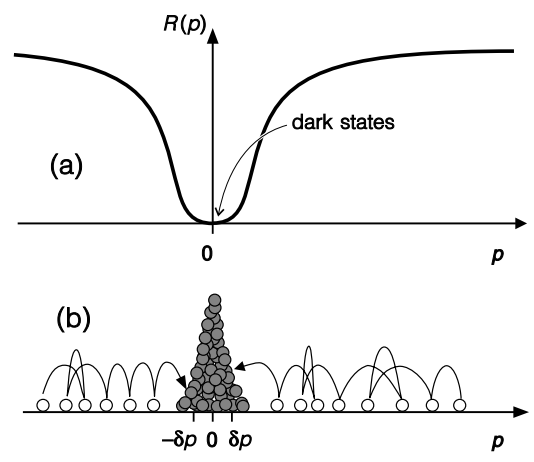}
        \end{overpic}
        \caption{Adapted from~\cite{Bardou2002}. L\'evy statistics and power-law distributions have meaningful consequences for other atomic systems. One such system is sub-recoil laser cooling. (top) The fluorescence rate $R(p)$ vanishes for $p \to 0$. (bottom) The atoms perform a random walk in momentum-space and accumulate in a small interval around $p = 0$ where they remain trapped.}
	{\label{fig:fig12}}
\end{figure}

Sub-recoil laser cooling is based on carefully engineering the photon scattering rate of an atomic ensemble in such a way that slow (cold) atoms have a smaller chance of absorbing a photon from the laser than hot ones. In other words, the photon scattering rate $R(p)\to0$ as the momentum $p\to0$. An atom will diffuse due to random kicks from photon recoil events until its momentum becomes small enough that the scattering rate diminishes significantly, and it will linger for a lengthy period of time at low momentum and remain cold (Fig.~\ref{fig:fig12}). The characteristic evolution time of an atom in such a laser field is $R^{-1}(p)$, which diverges as $p$ approaches zero. Depending on the small-$p$ behavior of the scattering rate the distribution of these waiting times may have heavy tails and even diverging moments - a signature of anomalous, L\'evy-type dynamics.

The rate $R(p)\sim |p|^\zeta$ and $\zeta$ can be controlled experimentally~\cite{Bardou1994,Reichel1995,cohen2011advances}, and one finds a non-normalized state which is controlled by the value of $\zeta$. In particular, sub-recoil laser cooling works best when the mean time the atom spends with momentum in a small interval close to zero diverges (thus states with momentum close to the minimum of $R(p)$ plotted in Fig.~\ref{fig:fig12} are extremely long-lived). This means, on the one hand, long stagnation times during which the speeds are close to zero, which is of course the goal of cooling, but at the same time indicates the violation of the basic postulates of statistical physics. 

The theoretical challenge is therefore to construct a statistical theory to replace ordinary ergodic theory, based on infinite ergodic theory~\cite{aaronson1997introduction,barkai2021transitions}. Consider the energy of the system, which in the absence of interactions is purely kinetic. Usually, for normal gases this is $k_B T/2$ per degree of freedom and one uses a perfectly normalized density to compute it, namely the Maxwell-Boltzmann momentum distribution. Here, one needs to use a non-normalized state with an important caveat: only when the kinetic energy is integrable with respect to the non-normalizable state, is this strategy valid. 

More specifically, just as for usual ensemble averages of physical observables, which are obtained by integrating over the steady-state densities, here the non-normalized state is used and if the result of the integration is finite, the observable is classified as ``integrable". This in turn means that observables like the energy of the system go through an ergodic transition as they switch from being integrable to non-integrable, similar to the Sisyphus cooling case described by Eq.~\ref{eqppp}, where the kinetic energy is calculated from a non-normalized state only when $1<\widetilde{U}_0 < 3$. However, for sub-recoil laser cooled gases, the infinite density is not a description of the large momentum tail of the distribution, but rather it is relevant when the system is in the coldest state possible. This is vastly different from the Sisyphus system, and for sub-Doppler cooling the anomalous statistics is found for shallow optical lattices, and far from the ideal cooling scheme.

\section{Discussion and Summary}
\label{Section:Discussion}

This Colloquium highlights the unique statistical properties of atoms undergoing ``Sisyphus cooling". The advantage of this system is twofold: its controllability, rarely found in experimental system exhibiting anomalous stochastic behavior, allows a single control parameter -- the depth of the optical lattice -- to control the various phases of the dynamics (Fig.~\ref{fig:fig10}). The analysis of the scale-free process is made possible with a relatively simple tool, a Fokker-Planck equation, without invoking fractional derivatives or other ad-hoc assumptions with regards to the power-law statistics of waiting times and jump distances. This relates to the second striking feature of this system which is its broad relevance. It was shown that the momentum distribution exhibits power-law statistics and a stable L\'evy distribution describes the spatial spreading of the packet of particles. Though such statistics appear in many systems, physically these power-laws cannot extend to infinity -- as, for example, the energy of the system is always bounded. 

We have shown how to employ the concept of infinite densities to describe the far tails of the corresponding densities. These non-normalized states are time-dependent solutions that match the heavy-tailed distributions and render finite otherwise infinite moments, like the energy or the mean-squared displacement. Treated for many years by mathematicians as a pure abstract theory, this infinite ergodic theory is here linked to an actual physical system. The relevance of these concepts extends beyond the cold atomic system and into the mathematical field of infinite ergodic theory, sub-recoil laser cooling (Sec.~\ref{Section:SubRecoil}), properties of $1/f$ noise~\cite{fox2021aging}, weak chaos, and stochastic renewal processes~\cite{Akimoto2020,Xu2022}.

In addition, and extending beyond the scope of this work, the moments of the spreading particles exhibit a bi-scaling behavior termed ``strong anomalous diffusion"~\cite{CASTIGLIONE199975,Aghion2017}. This means that the moments of some observable $o(t)$ obey $\braket{o^q}\sim t^{q\nu(q)}$ where $\nu(q)$ is a piece-wise linear function with a single jump in slope. Such behavior has been experimentally observed in the context of cellular dynamics in~\cite{Gal2010} and theorized to occur in systems such as hydrodynamics, infinite horizon Lorentz gasses and Sinai billiards.

We have seen how fundamental concepts, rooted deeply in our understanding of statistical physics, are violated in this system. Among those was Einstein-Green-Kubo formalism -- typically relating the diffusivity to a stationary correlation function -- which needed to be replaced with one that takes into account the ``aging" of the correlation function in this system. This results in the ability to calculate transport constants previously predicted to be infinite. This aging of correlation function is generally related to the $1/f$ noise spectrum -- for example in the context of protein diffusion on the cell membrane~\cite{fox2021aging} -- as well as to glassy systems, where magnetization correlations functions exhibit similar aging~\cite{Bouchaud1992}. Other fundamental implications reviewed were deviations from the Boltzmann-Gibbs equilibrium state and the equipartition theorem as well as the the application of infinite ergodic theory to exploration of the breakdown of ergodicity.

Looking forward to the missing pieces of the puzzle, another laser-cooling scheme emerges as both a theoretical and an experimental candidate for anomalous dynamics - that of Raman sideband cooling~\cite{Vuletic1998,Kerman2000,doi:10.1126/science.aan5614,zohar2022degenerate}. In this scheme, atoms are trapped in the potential wells of a far-detuned standing wave. The trapping beams induce Raman transitions and remove vibrational quanta, cooling the trapped atoms toward the lowest vibrational state. Optical pumping, needed to close the cooling cycle involves spontaneous Raman scattering that may  change the vibrational state and thus heat the atoms. For low vibrational states. such heating is minimized by the Lamb-Dicke effect~\cite{Vuletic1998}, yielding efficient cooling. However, at high vibrational states optical pumping results in excessive heating where non-standard statistics may become plausible. As it is typically preceded by pre-cooling into the Lamb-Dicke regime by other cooling methods, such anomalous statistics have not been observed so far. Equipartition violations in transient states have been observed and even used in the context of this scheme to optimize the cooling~\cite{doi:10.1126/science.aan5614,PhysRevResearch.2.023245}.

Another relevant aspect may be the effects of many-body physics on the anomalous statistics. How will atom-atom interactions drive the system to thermal equilibrium? Will this depend, on the depth of the optical lattice, and if so, how? Another unexplored aspect of the problem relates to the Green-Kubo formalism. In transport theory, the relation between the response to a linear external weak field, namely the calculation of the mobility of the system, is a standard problem. Further exploration into this issue could serve a wider audience interested in anomalous response functions. 

The study of the dynamics of the relaxation of the system to its steady-state, explored theoretically in~\cite{Hirschberg2011,Hirschberg2012}, requires temporal control over the lattice parameters - a feature inherently available in ultracold atomic experiments. By ``quenching" the power of the lattice lasers from different initial conditions to different final states one can explore how the system approaches the steady-state~\cite{Afek2019Thesis}. An ac temporal modulation of the lattice depth was performed in~\cite{Wickenbrock2012}. This renormalizes $\widetilde{U}_0$ and may allow access to an effective shallow lattice regime revealing the elusive Richardson phase of Eq.~\ref{eq:xMSD}~\cite{Barkai2014}.

Furthermore, one can consider two new experimental frontiers: the first statistics-focused and the second single-particle-focused, each with its own advantages. The former is rooted in the fact that rare events and heavy tails require many decades of signal-to-noise to be resolved properly, whereas the latter is exciting in the sense that direct observations of trajectories can yield insights into the underlying processes that are washed out in the measurements of a large ensemble of particles. In particular, with single particle trajectories one can in principle analyze the time averages computed from long trajectories, and see how they are related to ensemble averages. This will promote a better understanding of the ergodic hypothesis in systems with scale-free dynamics. Modification of the external confining potential is also expected to lead to intricate behaviors which have yet to be fully explored. A third, crossover, regime is now slowly being made accessible through recent advances in the trapping of large arrays of single atoms~\cite{Morgado2021} and ions~\cite{joshi2021observing} combining single-particle control with relatively large statistics.

\section{Acknowledgments}
The authors would like to thank Ariel Amir and Erez Aghion for their valuable input on the manuscript and Yoav Sagi, Andreas Dechant, Eric Lutz and Erez Aghion for important contributions to the work discussed in this Colloquium. The support of Israel Science Foundation's grant 1614/21 is acknowledged (DK, EB).

\bibliographystyle{apsrmp4-1}
\bibliography{AnomalousRMP}

\end{document}